\documentclass[a4paper,11pt]{article}
\usepackage{jheppub} 
\usepackage{amsfonts,amsmath} 
\usepackage{mathrsfs} 
\usepackage[dvipsnames]{xcolor}
\usepackage{mathtools}
\usepackage[T1]{fontenc}
\usepackage{amssymb,ascmac,fancybox,mathrsfs}
\usepackage{color,booktabs,fancyhdr,bm,comment,cancel}
\usepackage{hyperref,appendix}
\usepackage[]{graphicx,graphics}

\newcommand{\thalf}{{\tfrac{1}{2}}}
\newcommand{\half}{\frac{1}{2}}

\title{The effect of resummation on retarded Green's function and greybody factor in $AdS$ black holes}

\abstract{We investigate the retarded Green's function and the greybody factor in asymptotically AdS black holes. Using the connection coefficients of the Heun equation, expressed in terms of the Nekrasov-Shatashvili (NS) free energy of an $SU(2)$ supersymmetric gauge theory with four fundamental hypermultiplets, we derive asymptotic expansions for both the retarded Green's function and the greybody factor in the small horizon limit. Furthermore, we compute the corrections to  these asymptotic expansions resulting from the resummation procedure of the instanton part of the NS function.}

\author[a]{Juli\'{a}n Barrag\'{a}n Amado}
\emailAdd{jjamado@fc.ul.pt}
\affiliation[a]{Grupo de F\'{i}sica Matem\'{a}tica, Campo Grande Edif\'{i}cio C6,
Lisboa 1749--016, Portugal}

\author[b]{Shankhadeep
Chakrabortty,}
\emailAdd{s.chakrabortty@iitrpr.ac.in}
\affiliation[b]{Department of Physics, Indian Institute of Technology Ropar, Rupnagar, Punjab, India 140001.}

\author [b]{Arpit Maurya}
\emailAdd{arpit.20phz0009@iitrpr.ac.in }

\makeatletter
\renewcommand{\p@subsection}{}
\makeatother

\makeatletter
\newcommand{\xequal}[2][]{\ext@arrow 0055{\equalfill@}{#1}{#2}}
\def\equalfill@{\arrowfill@\Relbar\Relbar\Relbar}
\makeatother

\renewcommand{\thesection}{\arabic{section}}
\renewcommand{\thesubsection}{\arabic{section}.\arabic{subsection}}

\makeatletter
\renewcommand{\theequation}{\arabic{section}.\arabic{equation}}
\@addtoreset{equation}{section} 
\makeatother

\begin{document}
\count\footins = 1000 

\maketitle

\section{Introduction}
A recent resurgence of black hole perturbation theory using the techniques originally developed 
for the Nekrasov Shatashvili phase of the $\Omega$-background in 4D supersymmetric gauge theory \cite{Seiberg:1994rs, Seiberg:1994aj,Nekrasov:2002qd, Alday:2009aq, Nekrasov:2009rc} has opened up a significant possibility for achieving the exact analytical form of various physical observables 
including holographic thermal correlator, quasinormal modes, greybody factors and many more \cite{Hortacsu:2018rr,BarraganAmado:2023apy,He:2023wcs,Bonelli:2021uvf,Bianchi:2021mft, Bianchi:2021xpr,Consoli:2022eey,Lei:2023mqx, Bianchi:2023sfs, Fucito:2023afe,  Bautista:2023sdf}.
After the separation of variables, the equation of motion for the scalar mode in black hole perturbation theory results in linear second-order ordinary differential equations that can be represented in the form of Heun equations by appropriate redefinition of variables \cite{Aminov:2020yma}. Heun's equation is  known for its generality  as a second-order linear differential equation with four regular singular points \cite{ronveaux1995heun}. These singularities on the Riemann sphere characterize the equation as a generalization of the hypergeometric equation \cite{fiziev2015heun}.
The solution to Heun’s equation, also known as the Heun function near a singular point, can be expanded in terms of the solutions around other singularities. The connection coefficients specify the explicit form of such expansion and are computed in terms of the Nekrasov-Shatashvili partition function \cite {Aminov:2023jve}.

The connection coefficients and the underlying connection formula are best understood in the context of 2d Liouville CFT \cite{Bonelli:2022ten}. According to the 2D-4D duality of the Alday-Gaiotto-Tachikawa (AGT) correspondence, the BPS sector of four-dimensional $\mathcal{N} = 2, SU(2)$ gauge theory in $\Omega$ is dual to the Liouville CFT on the Riemann sphere \cite{Alday:2009aq}. This duality implies that the four-point function of the Liouville CFT corresponds to the Nekrasov-Shatashvili 
 partition function in $\mathcal{N} = 2, SU(2)$ gauge theory \cite{ Alday:2009fs}. In particular, the duality offers a fascinating connection between the semiclassical limit (large central charge limit) on the parameters of Liouville CFT and the Nekrasov-Shatashvili limit on the parameters of the $\Omega$ background in the $SU(2)$ gauge theory \cite{Nekrasov:2009rc}. The BPZ equation governing the five-point correlator in Liouville theory provides an avenue for series solutions through the use of conformal blocks. The interplay of conformal blocks in different OPE channels (s, t, and u channels) establishes connections among the solutions to the BPZ equation. In the limit of a large central charge (semi-classical limit), the BPZ equation transforms into Heun’s equation, a second-order linear differential equation with four regular singularities where these singular points signifies the insertion of primary operators in the theory. By the virtue of crossing symmetry, the relation among the solutions of the semi-classical BPZ equation leads to a connection formula among the solutions of Heun’s equation \cite{Bonelli:2022ten}.

Very recently, the authors of \cite{Dodelson:2022yvn}
established an exact analytical expression for the holographic retarded Green’s function for  AdS$_5$ Schwarzschild black hole, dual to the correlator in a thermal CFT living in the $R^1 \times S^3$ boundary of the black hole spacetime. 
They have used a connection formula between the Heun functions associated with the incoming mode expansion of a scalar field near the horizon and those near the boundary of the black hole spacetime. This connection formula nicely captures the response function and source terms present in the boundary expansion of the scalar mode. Finally, by following the prescription of Lorentzian Green's function of black hole spacetime \cite{Son:2002sd}, they obtained the retarded Green's function by taking the ratio of source to response function. Further, exact retarded Green’s function for a thermal CFT with chemical potential and angular momenta on $R^1 ×\times S^3$ and also for a thermal CFT living in $R^1 \times H^3$ are constructed in \cite{Bhatta:2022wga} and \cite{Bhatta:2023qcl} respectively.

The study of greybody factors in asymptotically flat spacetimes has been reviewed using the connection coefficients of the confluent Heun equation, resulting in asymptotic expansions proportional to the area of the black hole horizon in the low-energy limit \cite{Bonelli:2021uvf,BarraganAmado:2023apy}. Nevertheless, in the case of asymptotically AdS spcetimes, the role of the connection coefficients becomes more elusive due to the nature of the boundary at spatial infinity. Harmark and collaborators \cite{Harmark:2007jy} computed the greybody factors for static and spherically symmetric spacetime black holes in $d-$dimensions by splitting up the spacetime in three regions, such that the radial differential equation can be solved analytically, and then matched across the regions. Building on those considerations, \cite{Jorge:2014kra} and \cite{Noda:2022zgk} have investigated the greybody factors for rotating and non-asymptotically flat black holes.

In this work, we study the retarded Green's function and the greybody factor for 
$\text{Kerr-AdS}_5$ black hole and Reissner-Nordstr\"{o}m-AdS$_{\rm 5}$ black hole. In doing so, we analyze the uncharged and charged scalar mode perturbation in the respective black hole spacetimes. We show how the Klein-Gordon equation corresponding to the radial mode of scalar field takes the form of Heun's differential equation by successive applications of M\"{o}bius  transformation followed by a s-homotopic transformation. We find the analytic form of two independent solutions of the Heun's equation at each of the two regular singular points corresponding to the black hole horizon and the boundary of the black hole spacetime. We explicitly show the emergence of the connection formula and the connection coefficients and describe the role they play in the subsequent study of the retarded Green's function and greybody factor. As expected, the Nekrasov partition function plays a crucial role in determining the connection coefficients. Most interestingly, we show how the singularity appearing from the pole structure of the Nekrasov partition function is reincarnated in to the singularity of the pole structure of $a$, the vacuum expectation
value of the scalar in the vector hypermultiplets in the supersymmetric gauge theory, obtained by solving the Matone relation. Within the small black hole approximation, we use a resummation technique to cure the singularity structure in the $a$ order by order in radius of black hole horizon. This exercise surprisingly
accommodates a series correction terms in  $a$ that can be nicely presented in terms of the generating function of the Catalan numbers. Moreover, the presence of such correction terms in a closed form sigificantly simplify the final analytical results of Green's function and greybody factor.

The plan of the paper is
the following: In section \ref{sec:2} we discuss the explicit construction of the radial mode  of the scalar field in $\text{Kerr-AdS}_5$ black hole spacetime. In section \ref{sec:3} we describe the radial mode of a charged scalar field in Reissner-Nordstr\"{o}m-AdS$_{\rm 5}$ black hole spacetime.
Section \ref{sec:4} is dedicated to Heun's equation, the solutions to Heun's equation near regular singular point and the connection formula.
In section \ref{sec:5} and in section \ref{sec:6} we discuss the computtaions of retarded Green's function and Greybody factor for $\text{Kerr-AdS}_5$ and Reissner-Nordstr\"{o}m-AdS$_{\rm 5}$ black holes respetively.
Finally, we conclude in section \ref{sec:7}. 

\section{Scalar perturbations in Kerr-AdS$_{\mathrm{5}}$}
\label{sec:2}

In order to set the backdrop of a specific example of scalar perturbation theory in AdS black hole, let us review the five dimensional Kerr-${\rm AdS}_5$ black hole as presented in \cite{Hawking:1998kw}. The explicit form of the corresponding metric takes the form,
\begin{equation}\label{eq:metric}
\begin{split}
ds^{2} = &-\dfrac{\Delta_{r}}{\rho^{2}}\left(dt - \dfrac{\hat{a}_{1}\sin^{2}\theta}{\Xi_{1}}d\phi_{1} - \dfrac{\hat{a}_{2}\cos^{2}\theta}{\Xi_{2}}d\phi_{2}\right)^{2} + \dfrac{\Delta_{\theta}\sin^{2}\theta}{\rho^{2}}\left(\hat{a}_{1}dt - \dfrac{r^{2} + \hat{a}_{1}^{2}}{\Xi_{1}}d\phi_{1}\right)^{2} \\ 
&+ \dfrac{1 + r^{2}/L^{2}}{r^{2}\rho^{2}}\left(\hat{a}_{1}\hat{a}_{2}dt - \dfrac{\hat{a}_{2}(r^{2} + \hat{a}_1^{2})\sin^{2}\theta}{\Xi_{1}}d\phi_{1} - \dfrac{\hat{a}_{1}(r^{2} + \hat{a}_2^{2})\cos^{2}\theta}{\Xi_{2}}d\phi_{2} \right)^{2} \\ 
&+ \dfrac{\Delta_{\theta}\cos^{2}\theta}{\rho^{2}}\left(\hat{a}_{2}dt - \dfrac{r^{2} + \hat{a}_2^{2}}{\Xi_{2}}d\phi_{2}\right)^{2}
+ \dfrac{\rho^{2}}{\Delta_{r}}dr^{2} + \dfrac{\rho^{2}}{\Delta_{\theta}}d\theta^{2},
\end{split} 
\end{equation}
where
\begin{equation}\label{eq:delta}
\begin{gathered}
\Delta_{r} = \frac{1}{r^{2}}\left(r^{2} + \hat{a}_1^{2}\right)\left(r^{2} + \hat{a}_2^{2}\right)\left(1 + \frac{r^{2}}{L^{2}}\right) - 2M = \frac{1}{L^{2}r^{2}}\left(r^{2} - r_{0}^{2}\right)\left(r^{2} - r_{-}^{2}\right)\left(r^{2} - r_{+}^{2}\right), \\ 
\Delta_{\theta} = 1 - \frac{\hat{a}_{1}^{2}}{L^{2}}\cos^{2}\theta - \frac{\hat{a}_{2}^{2}}{L^{2}}\sin^{2}\theta, \qquad \ 
\rho^{2} = r^{2} + \hat{a}_1^{2}\cos^{2}\theta + \hat{a}_2^{2}\sin^{2}\theta, \\
\Xi_{1} = 1 - \frac{\hat{a}_{1}^{2}}{L^{2}}, \qquad \Xi_{2} = 1 - \frac{\hat{a}_{2}^{2}}{L^{2}}
\end{gathered}
\end{equation}
Here $L$ stands for AdS radius and $M$, $\hat{a}_{1}$ and $\hat{a}_{2}$ are the real parameters, related to the ADM mass (${\cal M}$) and angular momenta (${\cal J}_{1}, {\cal J}_{2}$) respectively \cite{Gibbons:2004ai,Hollands:2005wt} 
\begin{equation}
{\cal M} = \frac{\pi M(2\Xi_1+2\Xi_2-\Xi_1\Xi_2)}{4\Xi_1^2\Xi_2^2},\quad\quad
{\cal J}_{1} = \frac{\pi M \hat{a}_{1}}{2\Xi_1^2\Xi_2},\quad\quad
{\cal J}_{2} = \frac{\pi M \hat{a}_{2}}{2\Xi_1\Xi_2^2}.
\end{equation}
Within a physically sensible range of parameters described as $M>0$, and $\hat{a}^{2}_{1},\hat{a}^{2}_{2}<1$, 
$\Delta_r = 0$ allows two real roots $r_+$ and $r_-$ signifying the inner and outer horizons of the black hole respectively  and one purely imaginary root $r_0$ that satisfies the following relation \cite{Gibbons:2004ai},
\begin{equation}
  -r_{0}^{2} = L^{2} + \hat{a}_{1}^{2} + \hat{a}_{2}^{2} + r_{-}^{2}+r_{+}^{2}.
\end{equation}
For the purposes of this article, we will see the radial variable, or rather $r^2$, as a generic complex number. It will be interesting for us to treat all three roots of $\Delta_r$,e.g., $r_+^2,r_-^2$ and $r_0^2$ as Killing horizons. 
Actually, in the complexified version of the metric \eqref{eq:metric}, in all three hypersurfaces defined by $r=r_0,r_-$ and $r_+$ we have that each of the Killing fields \begin{equation}\label{eq:Killing}
\xi_{k} = \dfrac{\partial}{\partial t} +
\Omega_{1}(r_{k})\dfrac{\partial}{\partial \phi} +
\Omega_{2}(r_{k})\dfrac{\partial}{\partial \psi},\quad\quad k=0,-,+, 
\end{equation}
becomes null \cite{Frolov:1998wf}. 
The temperatures and angular velocities at each horizon are given by 
\begin{equation}\label{eq:temperature} 
\begin{gathered}
T_{k} = \frac{r^{2}_{k}\Delta_r'(r_{k})}{4 \pi (r^{2}_{k} + \hat{a}_{1}^{2})(r^{2}_{k} + \hat{a}_{2}^{2})} = \frac{r_{k}}{2 \pi L^{2}}\frac{(r_{k}^{2} - r_{i}^{2})(r_{k}^{2} - r_{j}^{2})}{(r_{k}^{2} + \hat{a}_{1}^{2})(r_{k}^{2} + \hat{a}_{2}^{2})},
\qquad  i \neq j\neq k, \\
\Omega_{k,1} = \dfrac{\hat{a}_{1} \Xi_{1}}{r^{2}_{k} + \hat{a}_{1}^{2}}, \qquad
\Omega_{k,2} = \dfrac{\hat{a}_{2} \Xi_{2}}{r^{2}_{k} + \hat{a}_{2}^{2}}.
\end{gathered}
\end{equation}
Within the physically sensible range of parameters, $T_+$ is positive, $T_-$ is negative and $T_0$ is purely imaginary.

\subsection{The Klein-Gordon equation}
\label{sec:2.1}

The Klein-Gordon (KG) equation for a scalar of mass $\mu$, in the background \eqref{eq:metric}, is determined by
\begin{equation}\label{eq:KG}
\frac{1}{\sqrt{-g}}\partial_{\mu}\left(\sqrt{-g} g^{\mu\nu} \partial_{\nu}\right)\Phi - \mu^{2}\Phi = 0.
\end{equation}
We consider the following ansatz
\begin{equation}
\label{kerransatz}
\Phi(t,r,\theta,\phi_{1},\phi_{2}) =  e^{-i\omega t + im_{1}\phi_{1} +im_{2}\phi_{2}}R(r)S(\theta),
\end{equation}
where $\omega$ is the frequency of the mode, and $m_{1}, m_{2} \in \mathbb{Z}$ are the azimuthal quantum numbers. With the suitable choice of ansatz \eqref{kerransatz},KG equation \eqref{eq:KG} decouples into two second-order ordinary differential equations of the form
\begin{subequations}
\begin{equation}\label{eq:radialode}
\begin{split}
\biggl[\dfrac{1}{r}\dfrac{d}{dr}\left(r\Delta_{r}\dfrac{d}{dr}\right) &+ \dfrac{(r^{2} + \hat{a}_{1}^{2})^{2}(r^{2} + \hat{a}_2^{2})^{2}}{r^{4}\Delta_{r}}\left(\omega - \dfrac{m_{1}\hat{a}_{1}\Xi_{1}}{r^{2} + \hat{a}_{1}^{2}} - \dfrac{m_{2}\hat{a}_{2}\Xi_{2}}{r^{2} + \hat{a}_{2}^{2}}\right)^{2} \\
&\qquad\qquad - \dfrac{1}{r^{2}}\left(\hat{a}_{1}\hat{a}_{2}\omega - \hat{a}_{2} \Xi_{1} m_{1} - \hat{a}_{1} \Xi_{2} m_{2}\right)^{2} - \lambda_{\ell} - \mu^{2}r^{2}\biggr]R(r) = 0,
\end{split}
\end{equation}
\begin{equation}\label{eq:angularode}
\begin{split}
\biggl[\dfrac{1}{\sin\theta\cos\theta}&\dfrac{d}{d\theta}\left(\sin\theta \cos\theta \Delta_{\theta}\dfrac{d}{d\theta}\right) 
- \omega^{2}L^{2} - \frac{m_{1}^{2}\Xi_{1}}{\sin^2\theta} - \frac{m_{2}^{2}\Xi_{2}}{\cos^2\theta} \\
&+ \frac{\Xi_{1}\Xi_{2}}{\Delta_{\theta}}\left(\omega L + m_1\frac{\hat{a}_{1}}{L} + m_{2}\frac{\hat{a}_2}{L}\right)^{2} 
- \mu^{2}\left(\hat{a}_{1}^{2}\cos^2\theta + \hat{a}_{2}^{2}\sin^2\theta\right) + \lambda_{\ell}\biggr]S(\theta) = 0,
\end{split}
\end{equation}
\end{subequations}
where $\lambda_{\ell}$ is the separation constant. We shall revisit the separation constant and describe its explicit form in section \ref{sec:5}. For computational convenience, we  introduce a dimensionless radial coordinate 
\begin{equation}
\tilde{r} = \frac{r}{L}, \qquad \tilde{r}_{k} = \frac{r_{k}}{L}, \qquad k = \lbrace 0, -, + \rbrace
\end{equation}
and as a result $\Delta_{\tilde{r}}$ takes the form as $\Delta_{\tilde{r}}=\frac{\Delta_{r}}{L^{2}}$. If we scale the remaining parameters as 
\begin{equation}\label{eq:scaled_param}
\tilde{\omega} = L\omega,  \qquad \tilde{a}_{i} = \frac{\hat{a}_{i}}{L} ,  \qquad \tilde{\mu} = L\mu,
\end{equation}
we obtain the dimensionless ODEs, which do not depend on the AdS radius\footnote{It can be seen that temperatures and angular velocities \eqref{eq:temperature} will scale as follows
\begin{equation}
T_{k} = \frac{\tilde{T}_{k}}{L}, \qquad \Omega_{k,1} = \frac{\tilde{\Omega}_{k,1}}{L}, \qquad \Omega_{k,2} = \frac{\tilde{\Omega}_{k,2}}{L}.
\end{equation}
}. In terms of the tilde notation, the radial equation \eqref{eq:radialode} takes the following form:

\begin{subequations}\label{eq:radialode_kads}
\begin{equation}
\begin{split}
\Biggl\lbrace \frac{d^{2}}{d \tilde{r}^{2}} + \biggl(\frac{1}{\tilde{r}} &+ \frac{\Delta^{\prime}_{\tilde{r}}}{\Delta_{\tilde{r}}}\biggr)\frac{d}{d \tilde{r}} + \dfrac{(\tilde{r}^{2}+\tilde{a}_1^{2})^{2}(\tilde{r}^{2}+\tilde{a}_2^{2})^{2}}{\tilde{r}^{4}\Delta^{2}_{\tilde{r}}}\left(\tilde{\omega} - \frac{m_{1}\tilde{a}_{1}\Xi_{1}}{\tilde{r}^{2}+\tilde{a}_1^{2}} - \frac{m_{2}\tilde{a}_{2}\Xi_{2}}{\tilde{r}^{2}+\tilde{a}_2^{2}}\right)^{2}\\
&\qquad\qquad - \frac{1}{\tilde{r}^{2}\Delta_{\tilde{r}}}\left(\tilde{a}_{1}\tilde{a}_{2}\tilde{\omega} - \tilde{a}_{2} \Xi_{1} m_{1} - \tilde{a}_{1} \Xi_{2} m_{2}\right)^{2} - \frac{\lambda_\ell}{\Delta_{\tilde{r}}} - \frac{\tilde{\mu}^{2}\tilde{r}^{2}}{\Delta_{\tilde{r}}}\Biggr\rbrace R(\tilde{r}) = 0,
\end{split}
\end{equation}
where
\begin{equation}
\Delta_{\tilde{r}} = \frac{\left(\tilde{r}^{2} - \tilde{r}_{0}^{2}\right)\left(\tilde{r}^{2} - \tilde{r}_{-}^{2}\right)\left(\tilde{r}^{2} - \tilde{r}_{+}^{2}\right)}{\tilde{r}^{2}}
\end{equation}
\end{subequations}
Equation \eqref{eq:radialode_kads} possesses four regular singular points in $\tilde{r}^2$ variable, located at the roots of $\Delta_{\tilde{r}}$ and the point at infinity.
The characteristic exponents of the Frobenius solutions near to each finite singularity are given by 
\begin{equation}
\beta_{k}^{\pm} = \pm\frac{1}{2}\theta_k, \quad k = \lbrace +,-,0 \rbrace 
\end{equation}
where
\begin{equation}\label{eq:thetas}
\theta_{k} = \dfrac{i}{2\pi}\left(\dfrac{\tilde{\omega} - m_{1}\tilde{\Omega}_{k,1} - m_{2}\tilde{\Omega}_{k,2}}{\tilde{T}_{k}}\right),
\end{equation}
and for $\tilde{r} = \infty$, we have
\begin{equation}\label{eq:thetas_inf}
\beta_{\infty}^{\pm} = \frac{1}{2}(2\pm\theta_\infty), \quad \theta_{\infty} = \sqrt{4 + \tilde{\mu}^{2}} \coloneq \Delta - 2.
\end{equation}
It turns out that $\theta_{+}$ is the variation of the entropy $\delta S$ of the black hole as it absorbs a quantum of frequency and angular momenta at the outer horizon. Furthermore, $\theta_{\infty}$ can be expressed in terms of the conformal dimension $\Delta$ of the scalar operator dual to a scalar field in the bulk with mass $\tilde{\mu}$, such that $\Delta(\Delta-4) = \tilde{\mu}^{2}$ \cite{Penedones:2016voo}.
To bring the radial equation \eqref{eq:radialode_kads} to the canonical Heun form, we perform a change of variables followed by a s-homotopic transformation\footnote{Note that, with this choice of variables, the solution at $z \to \infty$ will behave as $R(z) \sim z^{\pm\theta_{0}/2}$. },
\begin{equation}\label{eq:shomor}
z=\frac{\tilde{r}^2-\tilde{r}_-^2}{\tilde{r}^2-\tilde{r}_0^2}, \qquad R(z) = z^{\beta_{-}^{-}}(z_{0}-z)^{\beta_{+}^{-}}(1-z)^{\beta_{\infty}^{+}}f(z),
\end{equation}
where
\begin{equation}\label{eq:radialt0}
z_{0} = \frac{\tilde{r}_{+}^{2} - \tilde{r}_{-}^{2}}{\tilde{r}_{+}^{2} - \tilde{r}_{0}^{2}}.
\end{equation}
The equation for $f(z)$ is 
\begin{equation}\label{eq:heunradial}
\dfrac{d^{2}f}{dz^{2}} + \biggl[\dfrac{1 - \theta_{-}}{z} + \dfrac{1 - \theta_{+}}{z-z_0} + \dfrac{\Delta - 1}{z-1}\biggr]\dfrac{df}{dz} + \left( \frac{\kappa_1\kappa_2}{z(z-1)} - \frac{z_{0}(z_{0} - 1)K_0}{z(z - z_{0})(z - 1)}\right)f(z) =  0,
\end{equation}
where
\begin{subequations}
\begin{equation}
\kappa_{1} = \frac{1}{2}(\theta_{-}+\theta_{+}-\Delta-\theta_{0}), \qquad \kappa_{2} = \frac{1}{2}(\theta_{-}+\theta_{+}-\Delta+\theta_{0}),
\end{equation}
\begin{equation}\label{eq:accessorykradial}
\begin{split}
4z_{0}(z_{0} - 1)K_{0} = &-\frac{\left(\lambda_\ell + \Delta(\Delta-4)\tilde{r}_{-}^2 - \tilde{\omega}^{2}\right)}{\tilde{r}_{+}^{2} - \tilde{r}_{0}^{2}} - (z_{0} - 1)\left[(\theta_{-} + \theta_{+} - 1)^{2} - \theta_{0}^{2} - 1\right]\\
&- z_{0}\left[2(\theta_{+} - 1)(1 - \Delta) + (\Delta - 2)^{2} - 2\right],
\end{split}
\end{equation}
\end{subequations}
where $K_{0}$ is called the accessory parameter. We show the explicit form of angular differential equation in appendix \ref{appendix_B}.

\section{Scalar perturbations in Reissner-Nordstr\"{o}m-AdS$_{\rm 5}$}
\label{sec:3}

The line element of the Reissner-Nordstr\"{o}m-AdS$_{5}$ (RN-AdS$_{5}$) black hole is
\begin{equation}\label{eq:metric_RN}
ds^{2} = -f(r)dt^{2} + \frac{dr^{2}}{f(r)} + r^{2}d\Omega_{3}^{2}
\end{equation}
where $d\Omega_{3}^{2}$ is the metric of the unit three-sphere and the blackening function $f(r)$ reads
\begin{equation}
f(r) = 1 - \frac{M}{r^{2}} + \frac{Q^{2}}{r^{4}} + \frac{r^{2}}{L^{2}}
\end{equation}
where $M$, $Q$ and $L$ are related to the black hole ADM mass, charge, and the radius of AdS respectively. The function $f(r)$ is a polynomial of degree six with only even powers of $r$, such that the roots of $f(r)$ can be expressed as 
\begin{equation}
f(r)  = \frac{\left(r^{2} - r_{0}^{2}\right)\left(r^{2} - r_{1}^{2}\right)\left(r^{2} - r_{2}^{2} \right)}{L^{2}r^{4}}.
\end{equation}
Since we are interested in space-time configurations that possess black hole horizons, the outer horizon corresponds to the largest positive real root $r_{2} = r_{+}$, while the inner horizon is defined as $r_{1} = r_{-}$ satisfying $r_{-} \leq r_{+}$, and $r_{0}$ is purely imaginary for generic values of the mass and charge. Furthermore, one can write mass parameter in terms of the the radius of the outer horizon as
\begin{equation}
M = r_{+}^{2} + \frac{Q^{2}}{r_{+}^{2}} + \frac{r_{+}^{4}}{L^{2}},
\end{equation}
which implies
\begin{equation}
\begin{split}
&r_{-}^{2} = \frac{L^{2}}{2}\left(-1 - \frac{r_{+}^{2}}{L^{2}} + \sqrt{\left(1 + \frac{r_{+}^{2}}{L^{2}}\right)^{2} + \frac{4Q^{2}}{L^{2}r_{+}^{2}}}\right) \\
&r_{0}^{2} = \frac{L^{2}}{2}\left(-1 - \frac{r_{+}^{2}}{L^{2}} - \sqrt{\left(1 + \frac{r_{+}^{2}}{L^{2}}\right)^{2} + \frac{4Q^{2}}{L^{2}r_{+}^{2}}}\right)
\end{split}
\end{equation}
The temperature at each horizon is
\begin{equation}
T_{k} = \frac{1}{4\pi}\frac{d f(r)}{d r}\biggl\vert_{r=r_{k}} = \frac{1}{2\pi L^{2}}\frac{\left(r_{k}^{2} - r_{i}^{2}\right)\left(r_{k}^{2} - r_{j}^{2}\right)}{r_{k}^{3}}, \qquad i \neq j \neq k
\end{equation}
The electromagnetic potential of the charged black hole can be written as
\begin{equation}
A_{\mu}dx^{\mu} = \left(-\frac{\sqrt{3}}{2}\frac{Q}{r^{2}} + C\right)dt,
\end{equation}
and for a vanishing potential at spatial infinity, we set $C = 0$.

\subsection{The Klein-Gordon equation}
\label{sec:3.1}

In the charged black hole background \eqref{eq:metric_RN}, the Klein-Gordon equation for a massive charged scalar field perturbation reads as
\begin{equation}\label{eq:KG_rn}
\frac{1}{\sqrt{-g}}D_{\mu}\left(\sqrt{-g}g^{\mu\nu}D_{\nu}\right)\Phi - \mu^{2}\Phi = 0,
\end{equation}
with $D_{\mu} = \partial_{\mu} - i e A_{\mu}$, and $e$ and $\mu$ being the charge and the mass of the field, respectively. Equation \eqref{eq:KG_rn} can be decomposed into two second-order ODEs by implementing the following ansatz,
\begin{equation}
\Phi\left(t,r,\theta,\phi,\psi\right) =  e^{-i \omega t}Y^{m_{1},m_{2}}_{\ell}\left(\theta,\phi,\psi\right)R(r)
\end{equation}
The angular part reduces to the spherical harmonic function on the three-sphere $Y^{m_{1},m_{2}}_{\ell}\left(\theta,\phi,\psi\right)$, which satisfies the eigenvalue equation
\begin{equation}\label{eq:angular_eigeneqn}
\Delta Y^{m_{1},m_{2}}_{\ell}\left(\theta,\phi,\psi\right) = -\ell(\ell+2)Y^{m_{1},m_{2}}_{\ell}\left(\theta,\phi,\psi\right),
\end{equation}
where $\ell$ is the angular momentum quantum number, and $m_{1}$ and $m_{2}$ are integers associated with the magnetic quantum numbers. By means of \eqref{eq:angular_eigeneqn}, the differential equation for the radial function $R(r)$ takes the following form
\begin{equation}\label{eq:radialode_rn}
\left(\frac{1}{r}\frac{d}{d r}\left(r^{3}f(r)\frac{d}{d r}\right) + \frac{r^{2}}{f(r)}\left(\omega - \frac{\sqrt{3}}{2}\frac{e Q}{r^{2}}\right)^{2} - \mu^{2}r^{2} - \ell(\ell + 2)\right)R(r) = 0
\end{equation}
For numerical convenience, we define
\begin{equation}
\begin{split}
&\tilde{r} = \frac{r}{L}, \qquad \tilde{r}_{k} = \frac{r_{k}}{L}, \qquad f(\tilde{r}) = f(r),  \qquad k = \lbrace 0,-,+ \rbrace \\
&\tilde{Q} = \frac{Q}{L^{2}}, \qquad \tilde{\omega} = L\omega, \qquad \tilde{\mu} = L\mu, \qquad \tilde{e} = \frac{\sqrt{3}}{2} L e,
\end{split}
\end{equation}
so that equation \eqref{eq:radialode_rn} will take a dimensionless form:
\begin{subequations}\label{eq:radialode_tilde_rn} 
\begin{equation}
\left\lbrace\frac{d^{2}}{d\tilde{r}^{2}} + \left(\frac{3}{\tilde{r}} + \frac{f^{\prime}(\tilde{r})}{f(\tilde{r})}\right)\frac{d}{d\tilde{r}} + \frac{1}{f(\tilde{r})}\left[\frac{1}{f(\tilde{r})}\left(\tilde{\omega} - \frac{\tilde{e}\tilde{Q}}{\tilde{r}^{2}}\right)^{2} - \frac{\ell(\ell+2)}{\tilde{r}^{2}} - \tilde{\mu}^{2}\right]\right\rbrace R(\tilde{r}) = 0,
\end{equation}
where
\begin{equation}
f(\tilde{r}) = \frac{\left(\tilde{r}^{2} - \tilde{r}_{0}^{2}\right)\left(\tilde{r}^{2} - \tilde{r}_{-}^{2}\right)\left(\tilde{r}^{2} - \tilde{r}_{+}^{2}\right)}{\tilde{r}^{4}}.
\end{equation}
\end{subequations}
Equation \eqref{eq:radialode_tilde_rn} possesses four regular singular points in $\tilde{r}^2$ variable, located at the roots of $f(\tilde{r})$ and the point at infinity. The characteristic exponents of the Frobenius solutions near to each finite singularity are given by
\begin{equation}
\beta_{k}^{\pm} = \pm\frac{1}{2}\theta_k, \quad k = \lbrace +,-,0 \rbrace 
\end{equation}
where
\begin{equation}\label{eq:thetas_rn}
\theta_{k} = \dfrac{i}{2\pi \tilde{T}_{k}}\left(\tilde{\omega} - \frac{\tilde{e}\tilde{Q}}{\tilde{r}_{k}^{2}}\right),
\end{equation}
and for $\tilde{r} = \infty$, we have
\begin{equation}\label{eq:thetas_inf_rn}
\beta_{\infty}^{\pm} = \frac{1}{2}(2 \pm \theta_\infty), \quad \theta_{\infty} = \sqrt{4 + \tilde{\mu}^{2}} \coloneq \Delta - 2.
\end{equation}
Since our aim is to write the equation \eqref{eq:radialode_tilde_rn} in the canonical Heun form, we introduce a M\"{o}bius transformation followed by a s-homotopic transformation
\begin{equation}\label{eq:shomor_rn}
z=\frac{\tilde{r}^2-\tilde{r}_-^2}{\tilde{r}^2-\tilde{r}_0^2}, \qquad R(z) = z^{\beta_{-}^{-}}(z_{0}-z)^{\beta_{+}^{-}}(1-z)^{\beta_{\infty}^{+}}f(z),
\end{equation}
where
\begin{equation}\label{eq:radialt0_rn}
z_{0} = \frac{\tilde{r}_{+}^{2} - \tilde{r}_{-}^{2}}{\tilde{r}_{+}^{2} - \tilde{r}_{0}^{2}},
\end{equation}
that leads to an equation for $f(z)$
\begin{equation}\label{eq:heunradial_rn}
\dfrac{d^{2}f}{dz^{2}} + \biggl[\dfrac{1 - \theta_{-}}{z} + \dfrac{1 - \theta_{+}}{z-z_0} + \dfrac{\Delta - 1}{z-1}\biggr]\dfrac{df}{dz} + \left( \frac{\kappa_1\kappa_2}{z(z-1)} - \frac{z_{0}(z_{0} - 1)K_0}{z(z - z_{0})(z - 1)}\right)f(z) =  0,
\end{equation}
with
\begin{subequations}
\begin{equation}
\kappa_{1} = \frac{1}{2}(\theta_{-}+\theta_{+}-\Delta-\theta_{0}), \qquad \kappa_{2} = \frac{1}{2}(\theta_{-}+\theta_{+}-\Delta+\theta_{0}),
\end{equation}
\begin{equation}\label{eq:accessorykradial_rn}
\begin{split}
4z_{0}(z_{0} - 1)K_{0} = &-\frac{\ell(\ell+2) + \Delta(\Delta-4)\tilde{r}_{-}^2 - \tilde{\omega}^{2}}{\tilde{r}_{+}^{2} - \tilde{r}_{0}^{2}} - (z_{0} - 1)\left[(\theta_{-} + \theta_{+} - 1)^{2} - \theta_{0}^{2} - 1\right]\\
&- z_{0}\left[2(\theta_{+} - 1)(1 - \Delta) + (\Delta - 2)^{2} - 2\right]
\end{split}
\end{equation}
\end{subequations}
It is important to note that while the radial equation \eqref{eq:heunradial_rn} shares a similar formal structure with the radial equation \eqref{eq:heunradial}, they are not identical. The characteristic exponents in each case depend on the details of the background. The accessory parameter $K_{0}$ includes the separation constant introduced to decouple the angular and radial components, which, in the case of RN-AdS$_{5}$ reduces to the eigenvalues associated to the scalar spherical harmonics on the three-sphere \eqref{eq:angular_eigeneqn}. Conversely, the angular eigenvalue problem in Kerr-AdS$_{5}$ is more complex and it is tipically solved perturbatively in specific limits, e.g., see \cite{Cho:2011pb,PhysRevD.99.105006} for asymptotic expansions of $\lambda$ in the near-equally rotating limit.

\section{Heun functions and connection coefficients}
\label{sec:4}
In this section we show the explicit construction of the normal form of Heun’s equation. We elaborate upon the solution of the Heun’s equation near regular singular points and derive the connection formula which we frequently use in the computations of retarded Green’s
function and greybody factor in the subsequent sections.
 
The canonical Heun's differential equation reads
\begin{subequations}
\begin{equation}\label{eq:Heun_canonical}
\left(\frac{d^{2}}{d z^{2}} + \left(\frac{\gamma}{z} + \frac{\delta}{z - 1} + \frac{\epsilon}{z - t}\right)\frac{d}{d z} + \frac{\alpha\beta z - q}{z(z - t)(z - 1)}\right)f(z) = 0,
\end{equation}
where the coefficients satisfy the condition
\begin{equation}
\alpha + \beta + 1 = \gamma + \delta + \epsilon,
\end{equation}
\end{subequations}
and the complex modulus of $t$ is small, $\vert t \vert \ll 1$. Via the following transformation
\begin{equation}
f(z) = z^{-\gamma}(t - z)^{-\epsilon}(1 - z)^{-\delta}\psi(z),
\end{equation}
one can bring \eqref{eq:Heun_canonical} into its normal form
\begin{equation}\label{eq:normalHeun}
\left(\partial^{2}_{z} + \frac{\frac{1}{4} - a_{0}^{2}}{z^{2}} + \frac{\frac{1}{4} - a_{t}^{2}}{(z - t)^{2}} + \frac{\frac{1}{4} - a_{1}^{2}}{(z - 1)^{2}} - \frac{\frac{1}{2} - a_{0}^{2} - a_{t}^{2} - a_{1}^{2} + a_{\infty}^{2} + u}{z(z-1)} + \frac{u}{z(z - t)}\right)\psi(z) = 0,
\end{equation}
with the identification
\begin{equation}
\begin{split}
&a_{0} = \frac{1 - \gamma}{2}, \quad a_{t} = \frac{1 - \epsilon}{2}, \quad a_{1} = \frac{1 - \delta}{2}, \quad a_{\infty} = \frac{\alpha - \beta}{2},\\
&u  = \frac{\gamma\epsilon - 2q + 2t\alpha\beta - t(\gamma + \delta)\epsilon}{2(t - 1)},
\end{split}
\end{equation}
which can be written in terms of $a_{i}$'s and $u$ as
\begin{equation}
\begin{split}
&\alpha = 1 - a_{0} - a_{1} - a_{t} + a_{\infty}, \quad \beta = 1 - a_{0} - a_{1} - a_{t} - a_{\infty}, \\
&\gamma = 1 - 2a_{0}, \quad \delta = 1 - 2a_{1}, \quad \epsilon = 1 - 2a_{t}, \\
&q = \thalf + \left(a_{0}^{2} + a_{t}^{2} + a_{1}^{2} - a_{\infty}^{2}\right)t - a_{t} - a_{1}t + a_{0}\left(2a_{t} - 1 + t(2a_{1} - 1)\right) + \left(1 - t\right)u.
\end{split}
\end{equation}
By comparing with the radial ODEs \eqref{eq:heunradial} and \eqref{eq:heunradial_rn}, one can recognize that
\begin{subequations}
\begin{equation}
t = z_{0}
\end{equation}
\begin{equation}\label{eq:radial_dic}
a_{0} = \pm \frac{\theta_{-}}{2}, \qquad a_{t} = \pm \frac{\theta_{+}}{2}, \qquad a_{1} = \pm \frac{\theta_{\infty}}{2}, \qquad a_{\infty} = \pm \frac{\theta_{0}}{2},
\end{equation}
\begin{equation}\label{eq:u}
u = -z_{0} K_{0} + \frac{1}{2}\theta_{-}\left(1 - \theta_{+}\right) - \frac{1}{2}\Delta\left(1 - \theta_{+}\right) - \frac{\left(1 - \theta_{+}\right)\left(\Delta - 1\right)}{2\left(z_{0} - 1\right)}.
\end{equation}
\end{subequations}
Note that the explicit form of the  parameters, $\theta_{k}$, $\theta_{\infty}$, $z_{0}$ and $K_{0}$ depends on the individual details of the corresponding black hole solutions such as the Kerr-AdS$_{5}$  and the Reissner-Nordstr\"{o}m-AdS$_{5}$ black holes. 

In \cite{Bonelli:2021uvf,Bonelli:2022ten,Aminov:2023jve}, the authors provided explicit expressions for the local expansions of the Heun functions and their connection coefficients in terms of the Nekrasov partition functions. These concepts have been applied to study quasi-normal modes, tidal Love numbers and greybody factors in different space-times \cite{Bianchi:2021mft,Bianchi:2021xpr,Consoli:2022eey,Lei:2023mqx,Bianchi:2023sfs}, post-Newtonian (PN) dynamics in the two-body problem \cite{Fucito:2023afe,Fucito:2024wlg}, as well as the computation of  retarded Green's function in asymptotically AdS space-times \cite{Dodelson:2022yvn,Bhatta:2022wga,Bhatta:2023qcl}. 

Following \cite{Aminov:2023jve,Lei:2023mqx}, two linearly independent solutions around $z = t$ are given by
\begin{subequations}\label{eq:sols_heun}
\begin{equation}\label{eq:solt}
\begin{split}
f_{-}^{(t)}(z) = &\mathrm{HeunG}\left(\frac{t}{t - 1},\frac{q - t\alpha\beta}{1 - t},\alpha,\beta,\epsilon,\delta,\frac{z - t}{1 - t}\right) \\
f_{+}^{(t)}(z) = &(t - z)^{1 - \epsilon}\mathrm{HeunG}\biggl(\frac{t}{t - 1},\frac{q - \alpha\beta t}{1 - t} - (\epsilon - 1)\left(\gamma + \frac{\delta t}{t - 1}\right),\\
&\qquad\qquad\qquad\qquad \alpha - \epsilon + 1, \beta - \epsilon + 1,2 - \epsilon, \delta,\frac{z - t}{1 - t} \biggr),
\end{split}
\end{equation}
while the ones around $z = 1$ can be written as
\begin{equation}\label{eq:sol1}
\begin{split}
f_{-}^{(1)}(z) = &\left(\frac{z - t}{1 - t}\right)^{-\alpha}\mathrm{HeunG}\left(t,q + \alpha(\delta - \beta), \alpha,\delta + \gamma - \beta,\delta,\gamma,t\frac{1 - z}{t - z}\right)\\
f_{+}^{(1)}(z) = &(1 - z)^{1 - \delta}\left(\frac{z - t}{1 - t}\right)^{-\alpha - 1 + \delta}\mathrm{HeunG}\biggl(t,q - (\delta - 1)\gamma t - (\beta - 1)(\alpha - \delta + 1),\\
&\qquad\qquad\qquad\qquad\qquad -\beta + \gamma +1, \alpha - \delta + 1, 2 - \delta, \gamma,t\frac{1 - z}{t - z}\biggr).
\end{split}
\end{equation}
\end{subequations}

In addition, the connection formula between solutions near to $z = t$ and near to $z = 1$ is given by
\begin{equation}\label{eq:connection}
\begin{split}
&t^{-\frac{1}{2}+a_{0} \mp a_{t}}(1 - t)^{-\frac{1}{2} + a_{1}}e^{\mp \frac{1}{2}\partial_{a_{t}}F(t)}f_{\pm}^{(t)}(z) = \\
&\left(\sum_{\sigma = \pm}\mathcal{M}_{\pm \sigma}\left(a_{t},a;a_{0}\right)\mathcal{M}_{(-\sigma) -}\left( a,a_{1};a_{\infty} \right)t^{\sigma a}e^{-\frac{\sigma}{2}\partial_{a}F(t)}\right)(1 - t)^{-\frac{1}{2} + a_{t}}e^{ i\pi (a_{1} + a_{t})}e^{\frac{1}{2}\partial_{a_{1}}F(t)}f^{(1)}_{-}(z) +\\
&\left(\sum_{\sigma = \pm}\mathcal{M}_{\pm \sigma}\left(a_{t},a;a_{0}\right)\mathcal{M}_{(-\sigma) +}\left(a,a_{1};a_{\infty}\right)t^{\sigma a}e^{-\frac{\sigma}{2}\partial_{a}F(t)}\right)(1 - t)^{-\frac{1}{2} + a_{t}}e^{ i\pi (-a_{1} + a_{t})}e^{-\frac{1}{2}\partial_{a_{1}}F(t)}f^{(1)}_{+}(z),
\end{split}
\end{equation}
where $F(t)$ is the instanton part of the NS free energy and the connection coefficients $\mathcal{M}$'s are defined as
\begin{equation}
\mathcal{M}_{\theta\theta^{\prime}}\left(a_{1},a_{2};a_{3}\right) = \frac{\Gamma\left(-2\theta^{\prime}a_{2}\right)\Gamma\left(1 + 2\theta a_{1}\right)}{\Gamma\left(\frac{1}{2} + \theta a_{1} - \theta^{\prime}a_{2} + a_{3}\right)\Gamma\left(\frac{1}{2} + \theta a_{1} - \theta^{\prime}a_{2} - a_{3} \right)}.
\end{equation}

By combining \eqref{eq:normalHeun} with \eqref{eq:sols_heun} we construct the solution to the radial equations \eqref{eq:heunradial} and \eqref{eq:heunradial_rn}  at the regular singular point $z = z_{0}$ (or equivalently $z = t$) as  
\begin{equation}
f(z) = C_{z_{0}-}f_{-}^{(z_{0})}(z) + C_{z_{0}+}f_{+}^{(z_{0})}(z). 
\end{equation}
Consequently, the asymptotic behavior of the radial solution $R(z)$ at $z = z_{0}$ leads to 
\begin{equation}\label{eq:asymptotic_hor}
R(z) \simeq C_{z_{0}-}(z_{0} - z)^{-\theta_{+}/2} + C_{z_{0}+}(z_{0} - z)^{\theta_{+}/2}
\end{equation}
for which an incoming wave at the outer horizon $z = z_{0}\,(\tilde{r} = \tilde{r}_{+})$ requires that $C_{z_{0}+} = 0$. Then, the remaining radial solution according to the boundary condition will be given by 
\begin{equation}\label{eq:solhor}
R(z) = C_{z_{0}-}z^{-\theta_{-}/2}(z_{0} - z)^{-\theta_{+}/2}(1 - z)^{\Delta/2}f^{(z_{0})}_{-}(z).
\end{equation}
Now we can relate the solution around $z = z_{0}$ to the local solutions around $z = 1$ through the appropriate choice of the connection coefficients \eqref{eq:connection}. Namely, we have
\begin{equation}\label{eq:Cz01}
\begin{split}
f_{-}^{(z_{0})}(z) = &\biggl(\mathcal{M}_{--}\left(a_{t},a;a_{0}\right)\mathcal{M}_{+-}\left(a,a_{1};a_{\infty}\right) z_0^{-a} e^{\frac{1}{2}\partial_{a}F(z_0)} \\
&+ \mathcal{M}_{-+}\left(a_{t},a;a_{0}\right)\mathcal{M}_{--}\left(a,a_{1};a_{\infty}\right) z_0^{a} e^{-\frac{1}{2}\partial_{a}F(z_0)}\biggr) z_0^{\frac{1}{2} - a_{0} - a_{t}} \left(1 - z_0\right)^{a_{t}-a_{1}} \\
&\times e^{i \pi (a_{1} + a_{t})} e^{\frac{1}{2}\left(\partial_{a_{1}}F(z_0) - \partial_{a_{t}}F(z_0)\right)}f_{-}^{(1)}(z)\\
&+ \biggl(\mathcal{M}_{--}\left(a_{t},a;a_{0}\right)\mathcal{M}_{++}\left(a,a_{1};a_{\infty}\right) z_0^{-a} e^{\frac{1}{2}\partial_{a}F(z_0)} \\
&+ \mathcal{M}_{-+}\left(a_{t},a;a_{0}\right)\mathcal{M}_{-+}\left(a,a_{1};a_{\infty}\right) z_0^{a} e^{-\frac{1}{2}\partial_{a}F(z_0)}\biggr) z_0^{\frac{1}{2} - a_{0} - a_{t}} \left(1 - z_0\right)^{a_{t}-a_{1}} \\
&\times e^{i \pi (a_{t} - a_{1})} e^{-\frac{1}{2}\left(\partial_{a_{1}}F(z_0) + \partial_{a_{t}}F(z_0)\right)}f_{+}^{(1)}(z).
\end{split}
\end{equation}
 By introducing \eqref{eq:Cz01} into \eqref{eq:solhor}, we obtain the radial solution at the horizon $z = z_{0}$ in terms of two local solutions at $z = 1$ as follows
\begin{equation}\label{eq:solrad}
\begin{split}
R(z) &= C_{1-} C_{z_{0}-} z^{-\theta_{-}/2}(z_{0} - z)^{-\theta_{+}/2}(1 - z)^{\Delta/2}f^{(1)}_{-}(z) \\
&\qquad\qquad\qquad + C_{1+} C_{z_{0}-} z^{-\theta_{-}/2}(z_{0} - z)^{-\theta_{+}/2}(1 - z)^{\Delta/2}f^{(1)}_{+}(z),
\end{split}
\end{equation}
where $f^{(1)}_{\pm}(z)$ are given in \eqref{eq:sol1} and $C_{1,\pm}$ read
\begin{subequations}\label{eq:connection_coefs}
\begin{equation}
\begin{split}
C_{1-} &= \biggl(\mathcal{M}_{--}\left(a_{t},a;a_{0}\right)\mathcal{M}_{+-}\left(a,a_{1};a_{\infty}\right) z_{0}^{-a} e^{\frac{1}{2}\partial_{a}F(z_{0})} \\
&+ \mathcal{M}_{-+}\left(a_{t},a;a_{0}\right)\mathcal{M}_{--}\left(a,a_{1};a_{\infty}\right) z_{0}^{a} e^{-\frac{1}{2}\partial_{a}F(z_{0})}\biggr) z_{0}^{\frac{1}{2} - a_{0} - a_{t}} \left(1 - z_{0}\right)^{a_{t}-a_{1}} \\
&\times e^{i \pi (a_{1} + a_{t})} e^{\frac{1}{2}\left(\partial_{a_{1}}F(z_{0}) - \partial_{a_{t}}F(z_{0})\right)},
\end{split}
\end{equation}
\begin{equation}
\begin{split}
C_{1+} &= \biggl(\mathcal{M}_{--}\left(a_{t},a;a_{0}\right)\mathcal{M}_{++}\left(a,a_{1};a_{\infty}\right) z_{0}^{-a} e^{\frac{1}{2}\partial_{a}F(z_{0})} \\
&+ \mathcal{M}_{-+}\left(a_{t},a;a_{0}\right)\mathcal{M}_{-+}\left(a,a_{1};a_{\infty}\right) z_{0}^{a} e^{-\frac{1}{2}\partial_{a}F(z_{0})}\biggr) z_{0}^{\frac{1}{2} - a_{0} - a_{t}} \left(1 - z_{0}\right)^{a_{t}-a_{1}} \\
&\times e^{i \pi (a_{t} - a_{1})} e^{-\frac{1}{2}\left(\partial_{a_{1}}F(z_{0}) + \partial_{a_{t}}F(z_{0})\right)}.
\end{split}
\end{equation}
\end{subequations}
The asymptotic behavior of the radial solution \eqref{eq:solrad} corresponding to spatial infinity $\tilde{r} \rightarrow \infty \left(z \rightarrow 1\right)$
\begin{equation}\label{eq:solinf}
\begin{split}
R(r) \simeq C_{1-} C_{z_{0}-}\left(z_{0} - 1\right)^{-\frac{1}{2}\theta_{+}}&\left(\tilde{r}_{-}^{2} - \tilde{r}_{0}^{2}\right)^{\frac{1}{2}\Delta}\tilde{r}^{-\Delta} \\
&+ C_{1+} C_{z_{0}-}\left(z_{0} - 1\right)^{-\frac{1}{2}\theta_{+}}\left(\tilde{r}_{-}^{2} - \tilde{r}_{0}^{2}\right)^{\frac{1}{2}(4 - \Delta)}\tilde{r}^{\Delta - 4}
\end{split}
\end{equation}
where for $\Delta \geq 4$, the first term converges while the second one diverges at $\tilde{r} \rightarrow \infty$, and thus the asymptotic solutions correspond to normalizable and non-normalizable solutions, respectively. Furthermore, the retarded Green's function defined as the ratio of the response to the source yields 
\begin{subequations}\label{eq:propagator}
\begin{equation}\label{eq:Gretarded}
\begin{split}
G_{\rm ret}\left(\tilde{\omega},\lambda\right) &= \left(\tilde{r}_{-}^{2} - \tilde{r}_{0}^{2}\right)^{\Delta - 2}\frac{C_{1-}}{C_{1+}} \\
&= e^{\partial_{a_{1}}F(z_{0}) + i\pi 2a_{1}}\left(\tilde{r}_{-}^{2} - \tilde{r}_{0}^{2}\right)^{\Delta - 2} \frac{\Gamma (2a_{1})}{\Gamma (-2a_{1})}\frac{\Sigma_{1-}}{\Sigma_{1+}},
\end{split}
\end{equation}
where
\begin{equation}\label{eq:sigma_1m}
\begin{split}
\Sigma_{1-} &= \frac{e^{-\frac{1}{2}\partial_{a}F(z_{0})}\Gamma (1 - 2a)\Gamma (-2a)}{\Gamma\left(\frac{1}{2} - a - a_{0} - a_{t} \right) \Gamma\left(\frac{1}{2} - a + a_{0} - a_{t} \right) \Gamma\left(\frac{1}{2} - a + a_{1} - a_{\infty} \right) \Gamma\left(\frac{1}{2} - a + a_{1} + a_{\infty} \right)}\\
&\times\left(\frac{r_{+}^{2} - r_{-}^{2}}{r_{+}^{2} - r_{0}^{2}}\right)^{a}\\
&+ \frac{e^{\frac{1}{2}\partial_{a}F(z_{0})}\Gamma (1 + 2a)\Gamma (2a)}{\Gamma\left(\frac{1}{2} + a - a_{0} - a_{t} \right) \Gamma\left(\frac{1}{2} + a + a_{0} - a_{t} \right) \Gamma\left(\frac{1}{2} + a + a_{1} - a_{\infty} \right) \Gamma\left(\frac{1}{2} + a + a_{1} + a_{\infty} \right)}\\
&\times\left(\frac{r_{+}^{2} - r_{-}^{2}}{r_{+}^{2} - r_{0}^{2}}\right)^{-a}
\end{split}
\end{equation}
\begin{equation}\label{eq:sigma_1p}
\begin{split}
\Sigma_{1+} &= \frac{e^{-\frac{1}{2}\partial_{a}F(z_{0})}\Gamma (1 - 2a)\Gamma (-2a)}{\Gamma\left(\frac{1}{2} - a - a_{0} - a_{t} \right) \Gamma\left(\frac{1}{2} - a + a_{0} - a_{t} \right) \Gamma\left(\frac{1}{2} - a - a_{1} - a_{\infty} \right) \Gamma\left(\frac{1}{2} - a - a_{1} + a_{\infty} \right)}\\
&\times \left(\frac{r_{+}^{2} - r_{-}^{2}}{r_{+}^{2} - r_{0}^{2}}\right)^{a}\\
&+ \frac{e^{\frac{1}{2}\partial_{a}F(z_{0})}\Gamma (1 + 2a)\Gamma (2a)}{\Gamma\left(\frac{1}{2} + a - a_{0} - a_{t} \right) \Gamma\left(\frac{1}{2} + a + a_{0} - a_{t} \right) \Gamma\left(\frac{1}{2} + a - a_{1} - a_{\infty} \right) \Gamma\left(\frac{1}{2} + a - a_{1} + a_{\infty} \right)}\\
&\times\left(\frac{r_{+}^{2} - r_{-}^{2}}{r_{+}^{2} - r_{0}^{2}}\right)^{-a}.
\end{split}
\end{equation}
\end{subequations}
where the radial dictionary for $a_{i}, i = \lbrace 0,t,1,\infty \rbrace$ is defined in \eqref{eq:radial_dic}, and we will choose the one with the negative sign and we have verified that the final result is independent of this choice of sign. Expression \eqref{eq:Gretarded} is related to the propagator derived in \cite{Dodelson:2022yvn,Bhatta:2022wga,Bhatta:2023qcl} under the change $a_{1} \rightarrow -a_{1}$, as a result of taking into account the asymptotic analysis of the radial function $R(z)$, instead of $\psi(z)$.

\section{Small Kerr-AdS$_{5}$ black holes}
\label{sec:5}

In order to study the retarded Green's function and the greybody factor in asymptotically AdS$_{5}$ black holes, we will focus on the small radius limit, while considering equal rotation parameters $\tilde{a}_{1} = \tilde{a}_{2} = \tilde{a}$. For this special case the angular equation \eqref{eq:angularode} reduces to a hypergeometric differential equation with the angular eigenvalue given by 
\begin{equation}\label{eq:lambda}
\lambda_{\ell} = \left(1 - \tilde{a}^{2}\right)\left[\ell(\ell + 2) - 2\tilde{a}\tilde{\omega}(m_{1} + m_{2}) - \tilde{a}^{2}(m_{1}+m_{2})^{2}\right] + \tilde{a}^{2}\left(\tilde{\omega}^{2} + \Delta(\Delta - 4)\right).
\end{equation}
as it was investigated in \cite{Aliev:2008yk}. Furthermore, we can define a critical rotation parameter as the maximal rotation parameter at extremality $\left( T_{+} = 0\right)$
\begin{equation}
\tilde{a}_{c} = \tilde{r}_{+}\sqrt{1 + 2\tilde{r}_{+}^{2}},
\end{equation}
and we require $\tilde{a} \leq \tilde{a}_{c}$, in order to guarantee a regular outer horizon. We then parameterize the accessible values for the rotation parameter as $\tilde{a} = \alpha\,\tilde{a}_{c}$, where $\alpha \leq 1$ is a dimensionless extremality parameter for fixed $\tilde{r}_{+} > 0$. In the small $\tilde{r}_{+}$ limit, the indicial coefficients of the radial equation \eqref{eq:thetas} and \eqref{eq:thetas_inf}, as well as the conformal modulus \eqref{eq:radialt0} can be written as
\begin{subequations}
\begin{equation}
\begin{split}
&\theta_{-} = -\frac{i\left(m_{1} + m_{2}\right)\alpha}{1-\alpha^{2}} + \frac{i\alpha^{2}\left(1 + \alpha^{2}\right)\tilde{\omega}\tilde{r}_{+}}{1 - \alpha^{2}} - \frac{i\left(m_{1}+m_{2}\right)\alpha\left(1+\alpha^{2}\right)\left(1-3\alpha^{2}\right)\tilde{r}_{+}^{2}}{2\left(1-\alpha^{2}\right)}\\
&\qquad + \frac{i\alpha^{2}\left(3+5\alpha^{2}-7\alpha^{4}-5\alpha^{6}\right)\tilde{\omega}\tilde{r}_{+}^{3}}{2\left(1-\alpha^{2}\right)} + \frac{i\left(m_{1}+m_{2}\right)\alpha\left(5 + 36\alpha^{2} + 42\alpha^{4} - 52\alpha^{6} - 35\alpha^{8} \right)\tilde{r}_{+}^{4}}{8\left(1-\alpha^{2}\right)}\\ 
&\qquad + \mathcal{O}\left(\tilde{r}_{+}^{5}\right),
\end{split}
\end{equation}
\begin{equation}
\begin{split}
&\theta_{+} = -\frac{i\left(m_{1} + m_{2}\right)\alpha}{1-\alpha^{2}} + \frac{i\left(1 + \alpha^{2}\right)\tilde{\omega}\tilde{r}_{+}}{1 - \alpha^{2}} + \frac{i\left(m_{1} + m_{2}\right)\alpha\left(1+\alpha^{2}\right)\tilde{r}_{+}^{2}}{1 - \alpha^{2}} - \frac{2i\tilde{\omega}\tilde{r}_{+}^{3}}{1-\alpha^{2}}\\ 
&\qquad - \frac{i\left(m_{1}+m_{2}\right)\left(3-2\alpha^{2}\right)\tilde{r}_{+}^{4}}{2\left(1-\alpha^{2}\right)} + \mathcal{O}\left(\tilde{r}_{+}^{5}\right),
\end{split}
\end{equation}
\begin{equation}
\begin{split}
&\theta_{0} = \tilde{\omega} + \left(m_{1}+m_{2}\right)\alpha\tilde{r}_{+} - \frac{3}{2}\left(1+\alpha^{2}\right)^{2}\tilde{\omega}\tilde{r}_{+}^{2} - \frac{1}{2}\left(m_{1}+m_{2}\right)\alpha\left(3+10\alpha^{2}+5\alpha^{4}\right)\tilde{r}_{+}^{3}\\
&\qquad + \frac{1}{8}\left(23 + 28\alpha^{2} + 70\alpha^{4} + 100\alpha^{6} + 35\alpha^{8}\right)\tilde{\omega}\tilde{r}_{+}^{4} + \mathcal{O}\left(\tilde{r}_{+}^{5}\right),
\end{split}
\end{equation}
\begin{equation}
\theta_{\infty} = \Delta - 2,
\end{equation}
\begin{equation}
z_{0} = \left(1 - \alpha^{2}\right)\tilde{r}_{+}^{2} - 2\left(1 - \alpha^{2}\right)\left(1 + 2\alpha^{2} + 3\alpha^{4} + \alpha^{6}\right)\tilde{r}_{+}^{4} + \mathcal{O}\left(\tilde{r}_{+}^{6}\right),
\end{equation}
\end{subequations}
and the radial dictionary is defined as follows
\begin{equation}\label{eq:radial_dcitionary_kerr}
a_{0} = \pm\frac{\theta_{-}}{2}, \quad a_{t} = \pm\frac{\theta_{+}}{2}, \quad a_{1} = \pm\frac{\theta_{\infty}}{2}, \quad a_{\infty} = \pm\frac{\theta_{0}}{2}, \quad t = z_{0}.
\end{equation}

In addition, the Matone relation associates the accessory parameter of the Heun equation with the vacuum expectation value $a$ of the gauge theory \cite{Matone:1995rx}
\begin{equation}\label{eq:Matone}
u = - \frac{1}{4} - a^{2} + a_{0}^{2} + a_{t}^{2} + t\partial_{t}F(t),
\end{equation}
where $F(t)$ is the instanton part of the NS free energy defined in \eqref{eq:NS_function}. Then, assuming an expansion for small $\tilde{r}_{+}$ of the form 
\begin{equation}\label{eq:vev}
a = \sum_{n = 0}^{\infty}b_{n}\tilde{r}_{+}^{n}
\end{equation}
one can compute the coefficients of \eqref{eq:vev} recursively. The small $\tilde{r}_{+}$ expansion for $a$ reads
\begin{equation}\label{eq:small_a}
\begin{split}
&a = \frac{1}{2}\left(\ell+1\right) -\frac{\left(1+\alpha^{2}\right)^{2}\left(3\ell(\ell+2) - \Delta(\Delta-4) + 3\tilde{\omega}^{2}\right)}{8\left(\ell+1\right)}\tilde{r}_{+}^{2}\\ 
&\qquad - \frac{\left(m_{1}+m_{2}\right)\alpha\left(1+\alpha^{2}\right)^{2}\left(6\ell(\ell+2) + (\Delta-2)^{2} - \tilde{\omega}^{2}\right)\tilde{\omega}}{4\ell(\ell+1)(\ell+2)}\tilde{r}_{+}^{3} + \mathcal{O}\left(\tilde{r}_{+}^{4}\right)
\end{split}
\end{equation}
In addition, the derivatives of the NS free energy in the case of the equally rotating Kerr-AdS$_{5}$ black hole are
\begin{subequations}\label{eq:derivatives_of_F}
\begin{equation}
\partial_{a}F^{\rm inst} = -\frac{1}{2}\left(\ell+1\right)\left(1-\alpha^{4}\right)\tilde{r}_{+}^{2} + \frac{(m_{1}+m_{2})(\ell+1)\alpha\left(1+\alpha^{2}\right)^{2}\tilde{\omega}\left((\Delta-2)^{2} - \tilde{\omega}^{2}\right)}{\ell^{2}(\ell+2)^{2}}\tilde{r}_{+}^{3} + \mathcal{O}\left(\tilde{r}_{+}^{4}\right)
\end{equation}
\begin{equation}
\partial_{a_{1}}F^{\rm inst} = \frac{1}{2}\left(1-\alpha^{4}\right)\left(\Delta-2\right)\tilde{r}_{+}^{2} + \frac{(m_{1}+m_{2})\alpha\left(1+\alpha^{2}\right)^{2}(\Delta-2)\tilde{\omega}}{\ell(\ell+2)}\tilde{r}_{+}^{3} + \mathcal{O}\left(\tilde{r}_{+}^{4}\right)
\end{equation}
\begin{equation}
\begin{split}
&\partial_{a_{t}}F^{\rm inst} = -i\frac{(m_{1}+m_{2})\alpha\left(1+\alpha^{2}\right)\left(\ell(\ell+2) + (\Delta-2)^{2} - \tilde{\omega}^{2}\right)}{2\ell(\ell+2)}\tilde{r}_{+}^{2}\\ 
&\qquad + i(1+\alpha^{2})\left(\frac{\left(1+\alpha^{2}\right)\left(\ell(\ell+2) + (\Delta-2)^{2} -\tilde{\omega}^{2}\right)}{2\ell(\ell+2)} + \frac{(m_{1}+m_{2})^{2}\alpha}{\ell(\ell+2)}\right)\tilde{\omega}\tilde{r}_{+}^{3} + \mathcal{O}\left(\tilde{r}_{+}^{4}\right)
\end{split}
\end{equation}
\end{subequations}

Interestingly, $a$ and $\partial_{a}F^{\rm inst}$ are independent of the sign choice in \eqref{eq:radial_dcitionary_kerr}, while $\partial_{a_{1}}F^{\rm inst}$ and $\partial_{a_{t}}F^{\rm inst}$ will get a global minus sign depending on the choice. Moreover, the poles located at $a = \pm n/2$ for $n \in \mathbb{N}$ in the analytic expansion of the NS free energy \eqref{eq:NS_function} are translated into poles for the values of the angular momentum quantum number of the  form $\lbrace \ell, \ell-1, \ell-2, \ldots \rbrace$ and will appear in the higher order terms of the asymptotic expansions for \eqref{eq:small_a} and \eqref{eq:derivatives_of_F}. It has been pointed out that the resummation procedure can eliminate the existing poles in the instanton partition function \cite{Gorsky:2017ndg,Alekseev:2018kcn,Alekseev:2019gkl}.  Bearing this in mind, one of the authors computed the low-energy absorption cross section of a Reissner-Nordstr\"{o}m black hole in rainbow gravity by first applying the resummation procedure to the instanton contributions to the vacuum expectation value $a$, and then extending this method to the derivatives of the free energy \cite{BarraganAmado:2023apy}.

By inspecting the structure of $a$, one can recognize that up to a numerical factor, it coincides with the monodromy around two singular points in the Riemann-Hilbert map between the four-punctured Riemann sphere and Fuchsian systems \cite{PhysRevD.99.105006}. Therefore, one can attempt an ansatz for the s-wave case $(\ell = m_{1} = m_{2} = 0)$ of the form
\begin{equation}\label{eq:a_ell0}
a(\ell=0) = \frac{1}{2} - \nu_{0}\tilde{r}_{+}^{2} + \mathcal{O}\left(\tilde{r}_{+}^{3}\right)
\end{equation}
and replace into the Matone relation \eqref{eq:Matone}. The left-hand side is given by equation \eqref{eq:u}, with the radial accessory parameter $K_{0}$ taken from \eqref{eq:accessorykradial}, which for small $\tilde{r}_{+}$ yields
\begin{equation}\label{eq:lhs}
\begin{split}
\mathrm{lhs} = -\frac{1}{2} - \frac{(1+\alpha^{2})}{2}\biggl[(1 - \alpha^{2}) &+ \frac{\Delta(\Delta-4)}{2}\\
&- \frac{\tilde{\omega}^{2}}{2(1-\alpha^{2})^{2}} + \frac{\alpha^{2}(4 + \alpha^{2})\tilde{\omega}^{2}}{2(1-\alpha^{2})^{2}}\biggr]\tilde{r}_{+}^{2} + \mathcal{O}\left(\tilde{r}_{+}^{4}\right)
\end{split}
\end{equation}
while the right hand side organizes as follows
\begin{equation}\label{eq:rhs}
\begin{split}
\mathrm{rhs} = -\frac{1}{2} + \nu_{0}\tilde{r}_{+}^{2} &- \frac{\left(1+\alpha^{2}\right)^{2}(1+\alpha^{4})\tilde{\omega}^{2}}{4(1-\alpha^{2})^{2}}\tilde{r}_{+}^{2}+ \frac{1}{8}\left(1-\alpha^{4}\right)\left(\tilde{\omega}^{2} - (\Delta - 2)^{2}\right)\tilde{r}_{+}^{2}\\ 
&- 2\nu_{0}x\left(1 + x + 2x + 5x^{3} + 14x^{4} + \ldots\right)\tilde{r}_{+}^{2} + 
\mathcal{O}\left(\tilde{r}_{+}^{4}\right),
\end{split}
\end{equation}
where
\begin{subequations}
\begin{equation}\label{eq:Catalan}
1 + x + 2x^{2} + 5x^{3} + 14x^{4} + \ldots = \frac{1 - \sqrt{1 - 4x}}{2x},
\end{equation}
\begin{equation}\label{eq:sequence}
x = \frac{\left(1+\alpha^{2}\right)^{4}\tilde{\omega}^{2}\left((\Delta - 2)^{2} - \tilde{\omega}^{2}\right)}{2^{6}\nu_{0}^{2}}.
\end{equation}
\end{subequations}
The terms inside the parenthesis proportional to $\nu_0$ in equation \eqref{eq:rhs} are associated with the sequence of Catalan numbers. Consequently, one can introduce the generating function for these numbers to compute the first correction, $\nu_{0}$. Hence, equating \eqref{eq:lhs} and \eqref{eq:rhs} gives a quadratic equation for $\nu_{0}$ up to the order $\mathcal{O}\left(\tilde{r}_{+}^{2}\right)$:
\begin{equation}\label{eq:first_correction}
\frac{1}{8}\left[\left(1+\alpha^{2}\right)^{2}\left(\Delta(\Delta - 4) -  3\tilde{\omega}^{2}\right) + 8\nu_{0}\sqrt{1 + \frac{\left(1+\alpha^{2}\right)^{4}\tilde{\omega}^{2}\left(\tilde{\omega}^{2} - (\Delta - 2)^{2}\right)}{2^{4}\nu_{0}^{2}}}\right]\tilde{r}_{+}^{2} + \mathcal{O}\left(\tilde{r}_{+}^{4}\right) = 0.
\end{equation}
It turns out that $\nu_0$ in terms of the black hole parameters has a surprisingly simple form
\begin{equation}\label{eq:nu0}
\nu_{0} = \pm \frac{1}{8}\left(1 + \alpha^{2}\right)^{2}\sqrt{\left(3\tilde{\omega}^{2} - \Delta(\Delta - 4)\right)^{2} - 4\tilde{\omega}^{2}\left(\tilde{\omega}^{2} - (\Delta - 2)^{2}\right)},
\end{equation}
and $x$ introduced in
\eqref{eq:sequence} takes the following form,
\begin{equation}\label{eq:equis}
x = \frac{\tilde{\omega}^{2}\left((\Delta - 2)^{2} - \tilde{\omega}^{2}\right)}{\left(3\tilde{\omega}^{2} - \Delta(\Delta - 4)\right)^{2} - 4\tilde{\omega}^{2}\left(\tilde{\omega}^{2} - (\Delta - 2)^{2}\right)}.
\end{equation}
Notice that the series expansion for the vacuum expectation value \eqref{eq:a_ell0}, as well as the generating function of the Catalan numbers make sense if $\tilde{r}_{+}^{2} < \nu_{0} \leq 1$ and $\vert x \vert < 1/4$ for given $\tilde{\omega}$, $\alpha$ and $\Delta$. 

By the same token, one can apply the resummation procedure to the derivatives of the NS free energy as follows
\begin{subequations}\label{eq:derivatives_of_F_resum}
\begin{equation}\label{eq:dFda}
\partial_{a}F^{\rm inst} = - \left(2x + 3x^{2} + \frac{20}{3}x^{3} + \frac{35}{2}x^{4} + \frac{252}{5}x^{5} + \ldots\right) + \mathcal{O}\left(\tilde{r}_{+}^{2}\right)
\end{equation}
\begin{equation}\label{eq:dFda1}
\partial_{a_{1}}F^{\rm inst} = \frac{1}{2}\left(\Delta - 2\right)\left[\left(1 - \alpha^{4}\right)  + \frac{\left(1+\alpha^{2}\right)^{4}\tilde{\omega}^{2}}{4\nu_{0}}\left(1 + x + 2x^{2} + 5x^{3} + 14x^{4} + \ldots \right)\right]\tilde{r}_{+}^{2} + \mathcal{O}\left(\tilde{r}_{+}^{4}\right)
\end{equation}
\begin{equation}\label{eq:dFdat}
\partial_{a_{t}}F^{\rm inst} = -i\frac{\left(1+\alpha^{2}\right)^{2}\left(\tilde{\omega}^{2}-(\Delta-2)^{2}\right)\tilde{\omega}}{8\nu_{0}}\left(1 + x + 2x^{2} + 5x^{3} + 14x^{4} + \ldots\right)\tilde{r}_{+} + \mathcal{O}\left(\tilde{r}_{+}^{3}\right)
\end{equation}
\end{subequations}
where $\nu_{0}$ and $x$ are defined in \eqref{eq:nu0} and \eqref{eq:equis}, respectively. We have observed that higher order terms in $z_{0}$ from the analytic expansion of the instanton part of the NS free energy contribute to the lower order terms and consequently we get a series in $x$ as  the correction terms nicely re-summed in the form of the corresponding generating functions\footnote{
\begin{subequations}\label{eq:sequences}
\begin{equation}
1 + x + 2x^{2} + 5x^{3} + 14x^{4} + \ldots = \frac{1 - \sqrt{1 - 4x}}{2x},
\end{equation}
\begin{equation}
2x + 3x^{2} + \frac{20}{3}x^{3} + \frac{35}{2}x^{4} + \frac{252}{5}x^{5} + \ldots = \log 4 - 2\log\left(1 + \sqrt{1 - 4x}\right)
\end{equation}
\end{subequations}}

\begin{subequations}\label{eq:derivatives_of_F_with_x}
\begin{equation}\label{eq:dFda0}
\partial_{a}F^{\rm inst} \coloneq  \partial_{a}F^{\rm inst}_{(0)} +  \mathcal{O}\left(\tilde{r}_{+}^{2}\right) = \log 4 - 2\log\left(1 + \sqrt{1 - 4x}\right) +  \mathcal{O}\left(\tilde{r}_{+}^{2}\right) 
\end{equation}
\begin{equation}\label{eq:dFda12}
\begin{split}
\partial_{a_{1}}F^{\rm inst} &\coloneq  \partial_{a_{1}}F^{\rm inst}_{(2)}\tilde{r}_{+}^{2} +  \mathcal{O}\left(\tilde{r}_{+}^{4}\right)\\
&= \frac{1}{2}\left(\Delta - 2\right)\left[\left(1 - \alpha^{4}\right)  + \frac{\left(1+\alpha^{2}\right)^{4}\tilde{\omega}^{2}}{4\nu_{0}}\frac{1 - \sqrt{1 - 4x}}{2x}\right]\tilde{r}_{+}^{2} + \mathcal{O}\left(\tilde{r}_{+}^{4}\right)
\end{split}
\end{equation}
\begin{equation}\label{eq:dFdat1}
\partial_{a_{t}}F^{\rm inst} \coloneq  \partial_{a_{t}}F^{\rm inst}_{(1)}\tilde{r}_{+} +  \mathcal{O}\left(\tilde{r}_{+}^{3}\right) = -i\frac{\left(1+\alpha^{2}\right)^{2}\left(\tilde{\omega}^{2}-(\Delta-2)^{2}\right)\tilde{\omega}}{8\nu_{0}}\frac{1 - \sqrt{1 - 4x}}{2x}\tilde{r}_{+} + \mathcal{O}\left(\tilde{r}_{+}^{3}\right)
\end{equation}
\end{subequations}

\subsection{Retarded Green's function}
\label{sec:5.1}

In the previous section, we have computed the correction to $a(\ell=0)$, as well as the derivatives of the instanton part of the NS free energy, which appear in the connection coefficients of the solutions of the Heun equation \eqref{eq:connection}. Now we aim to derive an asymptotic expansion for the s-wave retarded Green's function in the equal angular momenta limit. This will be achieved by substituting \eqref{eq:nu0}, \eqref{eq:equis}, and \eqref{eq:derivatives_of_F_with_x} into \eqref{eq:Gretarded} and expanding for small $\tilde{r}_{+}$
\begin{equation}\label{eq:Gret0}
\begin{split}
&G_{\rm ret}\left(\tilde{\omega},\Delta\right) = \frac{e^{-i\pi\Delta}\Gamma\left(2-\Delta\right)\Gamma\left(\half(\Delta-\tilde{\omega}-2)\right)\Gamma\left(\half(\Delta+\tilde{\omega}-2)\right)}{\Gamma\left(\Delta-2\right)\Gamma\left(\half(2-\Delta-\tilde{\omega})\right)\Gamma\left(\half(2-\Delta+\tilde{\omega})\right)}\biggl\lbrace 1 + \biggl[\partial_{a_{1}}F^{\rm inst}_{(2)}\\ 
&+ \left(\Delta - 2\right)\left(1+2\alpha^{2}(1+\alpha^{2})\right) - \frac{4(\Delta-2)\nu_{0}}{\left((\Delta-2)^{2} - \tilde{\omega}^{2}\right)} +\frac{2\pi\sin \pi\Delta}{\cos \pi\tilde{\omega} - \cos \pi\Delta}\frac{e^{\partial_{a}F^{\rm inst}_{(0)}} - x}{e^{\partial_{a}F^{\rm inst}_{(0)}} + x}\nu_{0}\\
&+ \frac{3}{4}\left(1+\alpha^{2}\right)^{2}\tilde{\omega}\biggl(\psi^{(0)}\left(\thalf(\Delta - \tilde{\omega} - 2)\right) - \psi^{(0)}\left(\thalf(2 - \Delta - \tilde{\omega})\right) - \psi^{(0)}\left(\thalf(\Delta + \tilde{\omega} - 2)\right)\\
&+ \psi^{(0)}\left(\thalf(2 - \Delta + \tilde{\omega})\right)\biggr)\biggr]\tilde{r}_{+}^{2}\biggr\rbrace + \mathcal{O}\left(\tilde{r}_{+}^{3}\right),
\end{split}
\end{equation}
where $\psi^{(0)}(z)$ corresponds to the digamma function,  $\partial_{a}F^{\rm inst}_{(0)}$ and $\partial_{a_{1}}F^{\rm inst}_{(2)}$ are given by \eqref{eq:dFda0} and \eqref{eq:dFda12}, and refer to the coefficients of the series expansion in $\tilde{r}_{+}$ for the derivatives of the instanton part of the NS free energy with respect to $a$ and $a_{1}$, respectively. Note that expression \eqref{eq:Gret0} assumes that $\Delta$ is not an integer (see \cite{Jia:2024zes} regarding the case when $\Delta$ is an integer). Then, the asymptotic expansion for the retarded Green's function is given by
\begin{equation}\label{eq:Gret0_sum_kerrads5}
\begin{split}
&G_{\rm ret}\left(\tilde{\omega},\Delta\right) = \frac{e^{-i\pi\Delta}\Gamma\left(2-\Delta\right)\Gamma\left(\half(\Delta-\tilde{\omega}-2)\right)\Gamma\left(\half(\Delta+\tilde{\omega}-2)\right)}{\Gamma\left(\Delta-2\right)\Gamma\left(\half(2-\Delta-\tilde{\omega})\right)\Gamma\left(\half(2-\Delta+\tilde{\omega})\right)}\biggl\lbrace 1 + \frac{1}{4}\biggl[3\left(1+\alpha^{2}\right)^{2}\tilde{\omega}\\
& \times\biggl(\psi^{(0)}\left(\thalf(\Delta - \tilde{\omega} - 2)\right) - \psi^{(0)}\left(\thalf(2 - \Delta - \tilde{\omega})\right) - \psi^{(0)}\left(\thalf(\Delta + \tilde{\omega} - 2)\right) + \psi^{(0)}\left(\thalf(2 - \Delta + \tilde{\omega})\right)\biggr)\\
& - \left(1 + \alpha^{2}\right)^{2}\left(\frac{2(\Delta - 2)}{\left((\Delta - 2)^{2} - \tilde{\omega}^{2}\right)} + \frac{\pi\sin \pi\Delta}{\cos \pi\Delta - \cos \pi\tilde{\omega}} \right)\sqrt{\left(\Delta(\Delta - 4) - 3\tilde{\omega}^{2}\right)^{2}}\\  &+ 2\left(\Delta - 2\right)\left(3 + 4\alpha^{2} + 3\alpha^{4}\right)\biggr]\tilde{r}_{+}^{2}\biggr\rbrace + \mathcal{O}\left(\tilde{r}_{+}^{3}\right).
\end{split}
\end{equation}

\subsection{Greybody factor}
\label{sec:5.2}

The computation of the greybody factor in asymptotically AdS spacetimes is a bit more subtle than in asymptotically flat or dS spacetimes. Due to the nature of the boundary condition at spatial infinity, the radiation produced at the horizon can travel all the way to the spatial infinity to be reflected back towards the black hole. As a result, the thermal equilibrium of black holes in AdS is ensured by this infinite mechanism \cite{Harmark:2007jy}. At the level of the scattering problem, the required boundary conditions cannot be satisfied since one cannot identify an outgoing wave solution at infinity. Nevertheless, using an approximation scheme the greybody factors of static and spherically symmetric, and rotating black holes in asymptotically AdS have been studied in \cite{Harmark:2007jy,Jorge:2014kra}, respectively. By considering different regions of the spacetime, the radial equation simplifies and analytical solutions can be found. These solutions can be matched in an overlapping region, resulting in a consistent definition for the fluxes at the horizon and at spatial infinity.

In addition, \cite{Noda:2022zgk} has performed the numerical computation of the absorption cross section in a similar fashion way. The asymptotic analysis at infinity remains the same as \cite{Harmark:2007jy,Jorge:2014kra}, while in the near-horizon region, the radial solutions are given in terms of the Heun functions instead of the Hypergeometric functions. Our approach follows both ideas: an approximate radial equation in the far-region and Heun functions at the horizon. However, we will introduce the exact connection coefficients rather than the ratio of the Wronskians as \cite{Noda:2022zgk}. Finally, we will define the conserved fluxes as
\begin{equation}\label{eq:flux}
\mathcal{F} = \frac{1}{2i}\left(R^{*}\,\tilde{r}\Delta_{\tilde{r}}\,\frac{d R}{d\tilde{r}} - R\,\tilde{r}\Delta_{\tilde{r}}\,\frac{d R^{*}}{d\tilde{r}}\right),
\end{equation}
so that the greybody factor is the ratio between the flux at the horizon and the flux coming in from infinity
\begin{equation}\label{eq:gamma_ell}
\gamma^{(\ell)} = \frac{\mathcal{F}_{\rm hor}}{\mathcal{F}^{(\infty)}_{\rm in}}
\end{equation}

We consider radial equation in the far region approximation $\tilde{r} \gg 1$, such that the radial equation \eqref{eq:radialode_kads} reduces to
\begin{equation}\label{eq:radial_far}
R^{\prime\prime} + \frac{5}{\tilde{r}}R^{\prime}  - \left[\frac{\Delta(\Delta-4)}{\tilde{r}^{2}} + \frac{\ell(\ell+2)-\tilde{\omega}^{2}}{\tilde{r}^{4}}\right]R = 0,
\end{equation}
then, we introduce a new radial coordinate
\begin{equation}
u = \frac{\tilde{\omega}}{\tilde{r}}
\end{equation}
and consider the limit $u \ll \tilde{\omega}$. As a result, equation \eqref{eq:radial_far} yields
\begin{equation}
R^{\prime\prime} - \frac{3}{u}R^{\prime} + \left(1 - \frac{\ell(\ell+2)}{\tilde{\omega}^{2}} - \frac{\Delta(\Delta - 4)}{u^{2}}\right)R = 0,
\end{equation}
whose solution is given by the linear combination of the Bessel functions, $J_{\nu}(z)$ and $Y_{\nu}(z)$, of the form
\begin{equation}
R(u) = C_{1} u^{2} J_{\Delta - 2}\left(\sqrt{1-\tfrac{\ell(\ell+2)}{\tilde{\omega}^{2}}}u\right) + C_{2} u^{2} Y_{\Delta - 2}\left(\sqrt{1-\tfrac{\ell(\ell+2)}{\tilde{\omega}^{2}}}u\right),
\end{equation}
which can be written, more conveniently, in terms of the Hankel functions 
\begin{equation}\label{eq:hankel}
R(u) = \thalf (C_{1} - iC_{2}) u^{2} H_{\Delta-2}^{(1)}(u) + \thalf (C_{1} + iC_{2}) u^{2} H_{\Delta - 2}^{(2)}(u),
\end{equation}
since their asymptotic structures describe the incoming and outgoing part of the wave function. Namely, the $H_{\Delta-2}^{(1)}(u)$ is associated with the incoming part of the solution, while $H_{\Delta-2}^{(2)}(u)$ controls the outgoing part. In the limit $\tilde{r} \rightarrow \infty$ the radial solution for the s-wave case behaves like 
\begin{equation}\label{eq:asymptotic_hankel}
R(r) \simeq \left(C_{1} A_{\rm J} + C_{2} A_{\rm Y}\right)\tilde{r}^{-\Delta} + C_{2} B_{\rm Y} \tilde{r}^{\Delta - 4},
\end{equation}
where
\begin{equation}
\begin{gathered}
A_{\rm J} = \frac{4}{\Gamma(\Delta-1)}\left(\frac{\tilde{\omega}}{2}\right)^{\Delta}, \quad A_{\rm Y} = -\frac{4}{\pi}\cos\pi(\Delta-2)\Gamma(2-\Delta)\left(\frac{\tilde{\omega}}{2}\right)^{\Delta},\\ 
B_{\rm Y} = -\frac{4}{\pi}\Gamma(\Delta-2)\left(\frac{\tilde{\omega}}{2}\right)^{4-\Delta}.
\end{gathered}
\end{equation}
By comparing the asymptotic behavior of the solutions \eqref{eq:asymptotic_hankel} with \eqref{eq:solinf}, we obtain
\begin{equation}
C_{1} = \frac{A_{1-}}{A_{\rm J}}C_{1-} - \frac{A_{1+}A_{\rm Y}}{A_{\rm J}B_{\rm Y}}C_{1+}, \qquad C_{2} =\frac{A_{1+}}{B_{\rm Y}}C_{1+},
\end{equation}
where $C_{1\pm}$ are given in \eqref{eq:connection_coefs} and
\begin{equation}
A_{1-} = C_{z_{0}-}\left(z_{0} - 1\right)^{-\frac{1}{2}\theta_{+}}\left(\tilde{r}_{-}^{2} - \tilde{r}_{0}^{2}\right)^{\frac{1}{2}\Delta}, \quad A_{1+} = C_{z_{0}-}\left(z_{0} - 1\right)^{-\frac{1}{2}\theta_{+}}\left(\tilde{r}_{-}^{2} - \tilde{r}_{0}^{2}\right)^{\frac{1}{2}(4 - \Delta)}.
\end{equation}
Furthermore, by inspecting the radial solution \eqref{eq:hankel}, we have
\begin{equation}\label{eq:Cin_Cout}
C_{\rm in} \equiv \thalf (C_{1} - iC_{2}), \qquad C_{\rm out} \equiv \thalf (C_{1} + iC_{2}),
\end{equation}
such that one can reproduce the (incoming) outcoming coefficients in \cite{Noda:2022zgk}. Therefore, the asymptotic incoming radial solution with quantum numbers $\left(\ell=m_{1}=m_{2}=0\right)$ reads
\begin{equation}\label{eq:sol_in}
R^{(\infty)}_{\rm in}(\tilde{r}) = C_{\rm in} \left(\tfrac{\tilde{\omega}}{\tilde{r}}\right)^{2}H_{\Delta-2}^{(1)}\left(\tfrac{\tilde{\omega}}{\tilde{r}}\right),
\end{equation}
such that after substituting \eqref{eq:sol_in} into equation \eqref{eq:flux}, we get the incoming flux at spatial infinity
\begin{equation}\label{eq:flux_in}
\mathcal{F}^{(\infty)}_{\rm in} = - \frac{2 \vert C_{\rm in} \vert^{2} \tilde{\omega}^{4}}{\pi}
\end{equation}
On the other hand, the solution at the horizon is given by \eqref{eq:solhor}, and the associated flux at the horizon is
\begin{equation}\label{eq:flux_hor}
\begin{split}
\mathcal{F}_{\rm hor} = - 2^{\Delta}\vert C_{z_{0}-} \vert^{2} &\tilde{\omega} \tilde{r}_{+}^{3}\left(1+\tilde{r}_{+}^{2}\right)^{\Delta/2}\left(1+(1+2\tilde{r}_{+}^{2})\alpha^{2}\right)^{2}\\
&\left(\frac{\sqrt{1+\left(1+2(1+2\tilde{r}_{+}^{2})\alpha^{2}\right)^{2}\tilde{r}_{+}^{2}}}{1+\left(3+2(1+\tilde{r}_{+}^{2})\alpha^{2}\right)\tilde{r}_{+}^{2}+\sqrt{1+\tilde{r}_{+}^{2}}\sqrt{1+\left(1+2(1+2\tilde{r}_{+}^{2})\alpha^{2}\right)^{2}\tilde{r}_{+}^{2}}}\right)^{\Delta}
\end{split}
\end{equation}
The greybody factor given as the ratio of the flux at the horizon and the incoming flux at spatial infinity reads
\begin{equation}\label{eq:gamma0}
\begin{split}
\gamma^{(\ell = 0)} = \frac{\mathcal{F}_{\rm hor}}{\mathcal{F}^{(\infty)}_{\rm in}} = &\frac{\vert C_{z_{0}-}\vert^{2}}{\vert C_{\rm in} \vert^{2}}\frac{2^{\Delta - 1}\pi \tilde{r}_{+}^{3}}{\tilde{\omega}^{3}}\left(1+\tilde{r}_{+}^{2}\right)^{\Delta/2}\left(1+(1+2\tilde{r}_{+}^{2})\alpha^{2}\right)^{2}\\
&\qquad\left(\frac{\sqrt{1+\left(1+2(1+2\tilde{r}_{+}^{2})\alpha^{2}\right)^{2}\tilde{r}_{+}^{2}}}{1+\left(3+2(1+\tilde{r}_{+}^{2})\alpha^{2}\right)\tilde{r}_{+}^{2}+\sqrt{1+\tilde{r}_{+}^{2}}\sqrt{1+\left(1+2(1+2\tilde{r}_{+}^{2})\alpha^{2}\right)^{2}\tilde{r}_{+}^{2}}}\right)^{\Delta}
\end{split}
\end{equation}
Since we are interested in the contribution of the resummation technique for the greybody factor in the small $\tilde{r}_{+}$ limit, we will derive asymptotic expansions without and with the resummation procedure. For the former we replace \eqref{eq:small_a} and \eqref{eq:derivatives_of_F} into \eqref{eq:gamma0}, while keeping the expansions up to second order $\mathcal{O}\left(\tilde{r}_{+}^{2}\right)$ to avoid the pole at $\ell=0$. 
Bearing this in mind, the greybody factor reads
\begin{equation}\label{eq:gamma0_nosum_kerrads}
\begin{split}
\gamma^{(0)} = 2\pi^{2}\left(1+\alpha^{2}\right)^{2}\tilde{r}_{+}^{3}&\frac{2^{2\Delta + 6} \pi \tilde{\omega}^{2\Delta + 5}\sin^{2}\pi\Delta}{\left(\cos\pi\Delta - \cos\pi\tilde{\omega}\right)^{2}}\\
&\textcolor{purple}{\frac{\left(3\tilde{\omega}^{2} - \Delta(\Delta - 4)\right)^{4}}{\left((3\tilde{\omega}^{2} - \Delta(\Delta - 4))^{2} - \tilde{\omega}^{2}(\tilde{\omega}^{2} - (\Delta - 2)^{2})\right)^{2}}}
\frac{1}{\chi} + \mathcal{O}\left(\tilde{r}_{+}^{4}\right)
\end{split}
\end{equation}
where
\begin{equation}
\begin{split}
\chi = 4^{2\Delta}\Gamma\left(\thalf(\Delta - \tilde{\omega} - 2)\right)^{2}&\Gamma\left(\thalf(\Delta + \tilde{\omega} - 2)\right)^{2}\tilde{\omega}^{8}
+ 4^{4}\Gamma\left(\thalf(2 - \Delta - \tilde{\omega})\right)^{2}\Gamma\left(\thalf(2 - \Delta + \tilde{\omega})\right)^{2}\tilde{\omega}^{4\Delta}\\
&\qquad+ \frac{4^{\Delta + 4}\pi^{2}\tilde{\omega}^{2\Delta + 4}\cos 2\pi\Delta}{\left(\cos\pi\Delta - \cos\pi\tilde{\omega}\right)\left((\Delta - 2)^{2} - \tilde{\omega}^{2}\right)}.
\end{split}
\end{equation}
On the other hand, for the latter case we substitute \eqref{eq:nu0}, \eqref{eq:equis}, and \eqref{eq:derivatives_of_F_resum} into \eqref{eq:gamma0}, which implies
\begin{equation}\label{eq:gamma0_resum_kerrads}
\gamma^{(0)} = 2\pi^{2}\left(1+\alpha^{2}\right)^{2}\tilde{r}_{+}^{3}\frac{2^{2\Delta + 6} \pi \tilde{\omega}^{2\Delta + 5} \sin^{2}\pi\Delta}{\left(\cos\pi\Delta - \cos\pi\tilde{\omega}\right)^{2}}\textcolor{RoyalBlue}{\frac{e^{\partial_{a}F^{\rm inst}_{(0)}}}{\left(e^{\partial_{a}F^{\rm inst}_{(0)}} + x\right)^{2}}}
\frac{1}{\chi} + \mathcal{O}\left(\tilde{r}_{+}^{4}\right)
\end{equation}
We observe that the highlighted factor in \eqref{eq:gamma0_nosum_kerrads} is corrected by a factor (highlighted in blue) containing all the contributions given by the derivative of the instanton part of the free energy at zeroth order, $\partial_{a}F^{\rm inst}_{(0)}$. Furthermore, one can replace it by its generating function \eqref{eq:dFda0}, to obtain
\begin{equation}
\frac{e^{\partial_{a}F^{\rm inst}_{(0)}}}{\left(e^{\partial_{a}F^{\rm inst}_{(0)}} + x\right)^{2}} = 1,
\end{equation}
which reduces the greybody factor \eqref{eq:gamma0_resum_kerrads} to
\begin{equation}\label{eq:gamma0_resum_kerrads5}
\gamma^{(0)} = 2^{2\Delta + 6} \pi \tilde{\omega}^{2\Delta + 5}\tilde{A}\frac{\sin^{2}\pi\Delta}{\left(\cos\pi\Delta - \cos\pi\tilde{\omega}\right)^{2}}\frac{1}{\chi} + \mathcal{O}\left(\tilde{r}_{+}^{4}\right),
\end{equation}
where $\Delta$ is not integer, and $\tilde{A} = 2\pi^{2}\left(1+\alpha^{2}\right)^{2}\tilde{r}_{+}^{3}$ is related to the area of the five-dimensional Kerr-AdS black hole with equal angular momenta in the small-radius limit. Note that in the non-rotating limit $\alpha \to 0$, equation \eqref{eq:gamma0_resum_kerrads5} gives the greybody factor for massive scalar fields in small Schwarzschild-AdS$_{5}$ black holes.

\begin{figure}[h]
\centering
\includegraphics[width=0.6\linewidth]{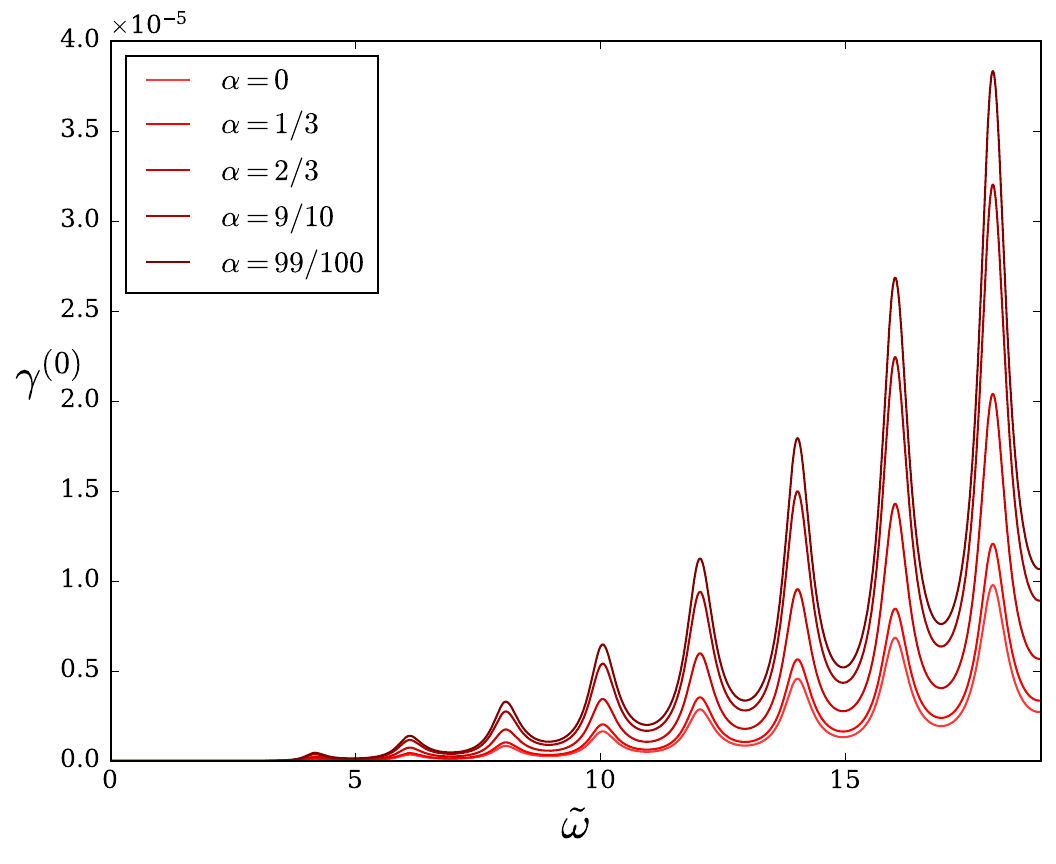}
\caption{Greybody factor \eqref{eq:gamma0_resum_kerrads5} as a function of $\tilde{\omega}$ for different values of $\alpha$, and fixed $\tilde{r}_{+} = 1/1000$, and $\Delta = 41/10$.}
\label{fig:1}
\end{figure}

\begin{figure}[h]
\centering
\includegraphics[width=0.9\linewidth]{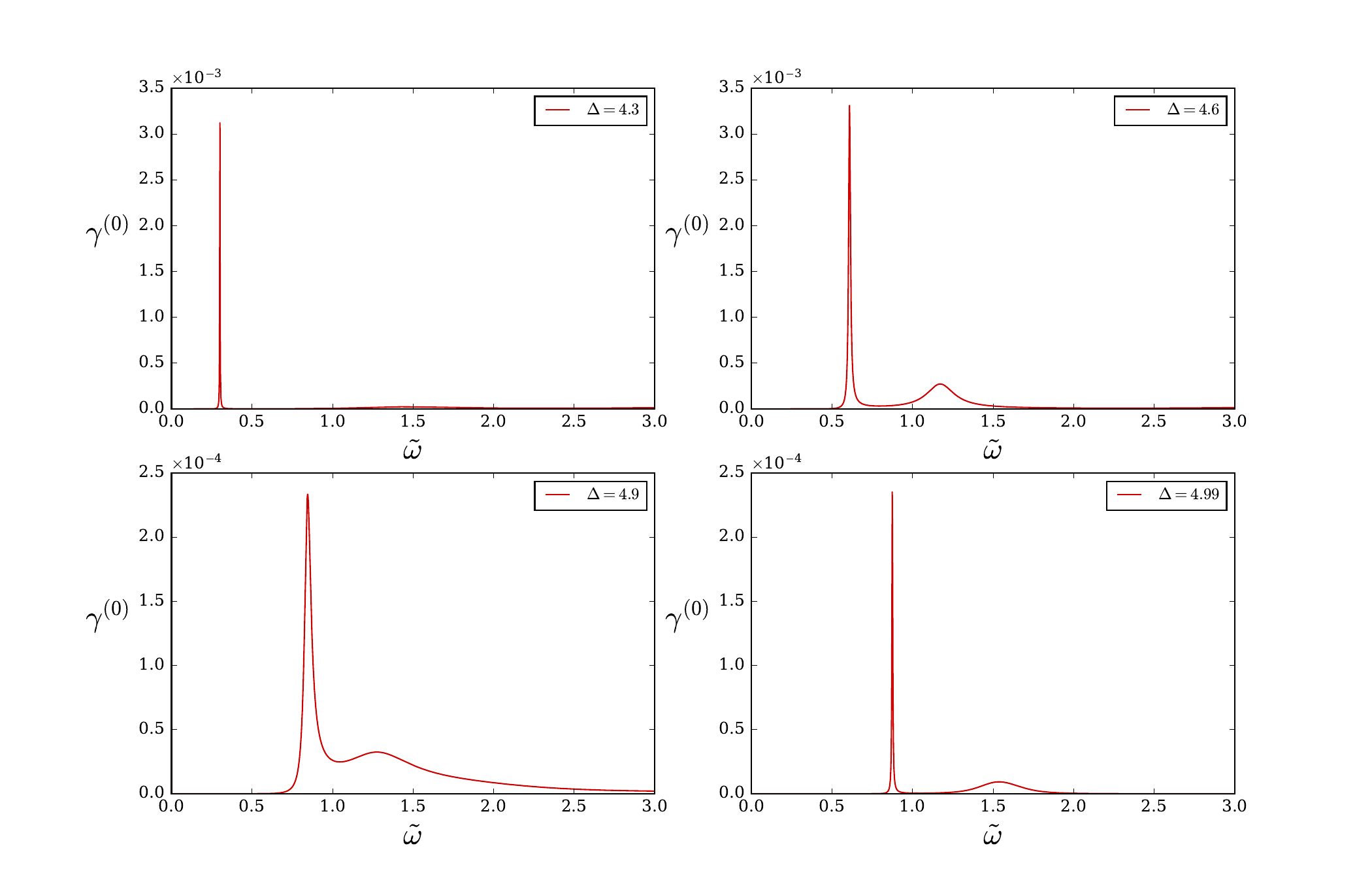}
\caption{Greybody factor as a function of $\tilde{\omega}$ for fixed values of $\alpha$ and $\tilde{r}_{+}$, while varying $\Delta$.}
\label{fig:2}
\end{figure}

In Fig. \ref{fig:1}, we present the greybody factor for small Kerr-AdS$_{5}$ black holes with equal rotation parameters and $\left(\ell = m_{1} = m_{2} = 0\right)$ modes, as calculated using formula \eqref{eq:gamma0_resum_kerrads5}. The analysis is performed for different values of the extremality parameter $\alpha = \lbrace 0, 1/3, 2/3, 9/10, \newline 99/100 \rbrace$, while keeping fixed $\tilde{r}_{+} = 1/1000$, and $\Delta = 41/10$, and varying the frequency $\tilde{\omega}$. It is worth mentioning that the greybody spectrum increases with $\alpha$ for fixed $\tilde{\omega}$, which contrasts with previous results for massless scalar fields in rotating cohomogeneity$-1$ BH space-times \cite{Jorge:2014kra}. As observed in \cite{Harmark:2007jy}, the large-amplitude oscillations in the greybody spectrum are consistent with the spacing between the normal modes frequencies in pure five-dimensional AdS spacetime. In addition, we observed an intriguing dynamics in the spectrum at small $\tilde{\omega}$. As we increase the conformal dimension $\Delta$, two peaks appear and start to move closer, merging and then separating again, as illustrated in Fig. \ref{fig:2}.

\begin{figure}[h]
\centering
\includegraphics[width=0.6\linewidth]{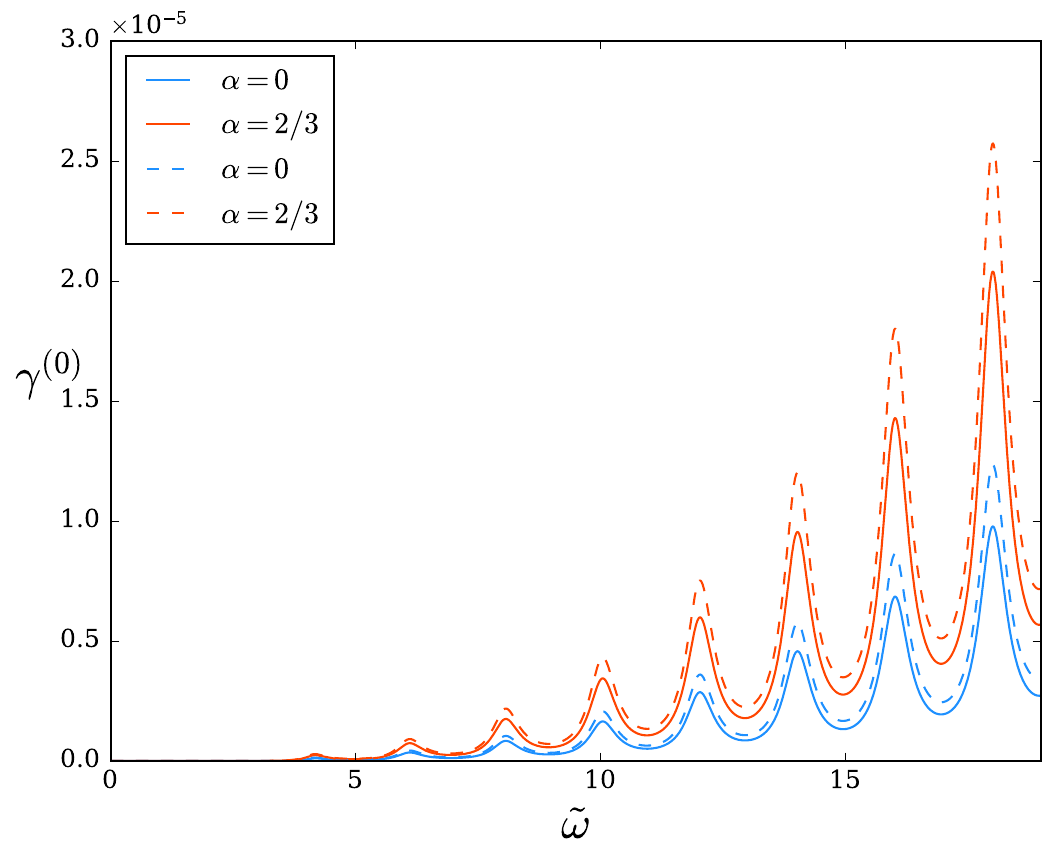}
\caption{Comparison between the greybody factors with (solid lines) and without resummation (dashed lines) for different values of $\alpha$ and fixed $\tilde{r}_{+} = 1/1000$, and $\Delta = 41/10$.}
\label{fig:3}
\end{figure}

In Fig. \ref{fig:3}, we compare the greybody factor calculated with and without the resummation procedure. The solid curves, representing the asymptotic formula with resummation \eqref{eq:gamma0_resum_kerrads5}, display a compressed spectrum compared to the dashed curves, which correspond to the expression without resummation \eqref{eq:gamma0_nosum_kerrads}.

\section{Small Reissner-Nordstr\"{o}m-AdS$_{\rm 5}$ black holes}
\label{sec:6}
We now turn to study the retarded Green's function and the greybody factor of a RN-AdS$_{5}$ black hole in the small-radius limit. To this matter, it is convenient to express the charge $\tilde{Q}$ in terms of $\tilde{r}_{+}$. Specifically, the temperature at the outer horizon defined as
\begin{equation}\label{eq:temp_rn}
\tilde{T}_{+} = \frac{1}{2\pi}\left[\frac{1}{\tilde{r}_{+}} - \frac{\tilde{Q}^{2}}{\tilde{r}_{+}^{5}} + 2\tilde{r}_{+}\right],
\end{equation}
vanishes at extremality, indicating that the maximal charge is given by
\begin{equation}
\tilde{Q}_{c} = \tilde{r}_{+}^{2}\sqrt{1 + 2\tilde{r}_{+}^{2}},
\end{equation}
such that $Q \leq Q_{c}$ must be satisfied to ensure a regular outer horizon. We then parametrize the accessible charge values as
\begin{equation}\label{eq:charge}
\tilde{Q} = q\,\tilde{Q}_{c} = q\,\tilde{r}_{+}^{2}\sqrt{1 + 2\tilde{r}_{+}^{2}}, \qquad 0 \leq q \leq 1,
\end{equation}
where $q$ is an extremality parameter, which describes  the Schwarzschild-AdS$_{5}$ black-hole solution for $q=0$ and the extremal RN-AdS$_{5}$ black hole for $q=1$ . Plugging \eqref{eq:charge} into \eqref{eq:thetas_rn} and expanding for small $\tilde{r}_{+}$ yields
\begin{subequations}\label{eq:thetas_and_z0_rn}
\begin{equation}
\theta_{-} = \frac{i\left(q\tilde{\omega} - \tilde{e}\right)q^{2}\tilde{r}_{+}}{1 - q^{2}} - \frac{i\left(\tilde{e} - 3q^{2}\tilde{e} - q\left(1 - 5q^{2} \right)\tilde{\omega}\right)q^{2}\tilde{r}_{+}^{3}}{2\left(1 - q^{2}\right)} + \mathcal{O}\left(\tilde{r}_{+}^{5}\right),
\end{equation}
\begin{equation}
\theta_{+} = \frac{i\left(\tilde{\omega} - q\tilde{e}\right)\tilde{r}_{+}}{1-q^{2}} - \frac{i\left(2\tilde{\omega} - q\tilde{e}\right)\tilde{r}_{+}^{3}}{1 - q^{2}} + \mathcal{O}\left(\tilde{r}_{+}^{5}\right),
\end{equation}
\begin{equation}
\theta_{0} = \tilde{\omega} - \left(\frac{3}{2}\left(1 + q^{2}\right)\tilde{\omega} - q\tilde{e}\right)\tilde{r}_{+}^{2} + \frac{1}{8}\left(\left(23 + 26q^{2} + 35q^{4}\right)\tilde{\omega} - 4\left(3 + 5q^{2}\right)q\tilde{e}\right)\tilde{r}_{+}^{4} + \mathcal{O}\left(\tilde{r}_{+}^{6}\right),
\end{equation}
\begin{equation}
\theta_{\infty} = \Delta - 2,
\end{equation}
while the conformal modulus \eqref{eq:radialt0_rn} reads
\begin{equation}
z_{0} = \left(1 - q^{2}\right)\tilde{r}_{+}^{2} - 2\left(1 - q^{4}\right)\tilde{r}_{+}^{4} + \mathcal{O}\left(\tilde{r}_{+}^{6}\right)
\end{equation}
\end{subequations}
Next, one can solve Matone's relation \eqref{eq:Matone} to obtain an expansion for the vacuum expectation value $a$ in the form
\begin{equation}\label{eq:small_a_rn}
a = \frac{1}{2}(\ell + 1) - \left(\frac{\left(1 + q^{2}\right)\left(3\tilde{\omega}^{2} - \Delta(\Delta - 4) + 3\ell(\ell + 2)\right)}{8(\ell + 1)} - \frac{q\tilde{e}\tilde{\omega}}{2(\ell + 1)}\right)\tilde{r}_{+}^{2} + \mathcal{O}\left(\tilde{r}_{+}^{4}\right) 
\end{equation}
analogous to \eqref{eq:small_a}. In addition, the derivatives of the NS free energy are
\begin{subequations}\label{eq:derivatives_rn}
\begin{equation}
\partial_{a}F^{\rm inst} = -\frac{1}{2}\left(1-q^{2}\right)\left(\ell + 1\right)\tilde{r}_{+}^{2} + \mathcal{O}\left(\tilde{r}_{+}^{4}\right)
\end{equation}
\begin{equation}
\partial_{a_{1}}F^{\rm inst} = \frac{1}{2}\left(1 - q^{2}\right)\left(\Delta - 2\right)\tilde{r}_{+}^{2} + \mathcal{O}\left(\tilde{r}_{+}^{4}\right)
\end{equation}
\begin{equation}
\partial_{a_{t}}F^{\rm inst} = \frac{i\left(\tilde{\omega} - q\tilde{e}\right)\left((\Delta - 2)^{2} + \ell(\ell + 2) - \tilde{\omega}^{2}\right)\tilde{r}_{+}^{3}}{2\ell(\ell + 2)} + \mathcal{O}\left(\tilde{r}_{+}^{5}\right)
\end{equation}
\end{subequations}
Due to the presence of a pole structure in the angular momentum quantum number $\ell$ within the expansions \eqref{eq:small_a_rn} and \eqref{eq:derivatives_rn}, it is necessary to sum the contributions from all orders in $z_{0}$, as demonstrated in \cite{Amado:2021erf}. Specifically, we limit our analysis to the s-wave case $(\ell = 0)$, following the same procedure introduced for the equal angular momenta Kerr-AdS$_{5}$ black hole in section \ref{sec:5}. In other words, we consider an ansatz of the form
\begin{equation}\label{eq:a0_rnads}
a(\ell=0) = \frac{1}{2} - \nu_{0}\tilde{r}_{+}^{2} + \mathcal{O}\left(\tilde{r}_{+}^{4}\right)
\end{equation}
and replace into the Matone's relation \eqref{eq:Matone}. The left-hand side is given by equation \eqref{eq:u}, with the radial accessory parameter $K_{0}$ taken from \eqref{eq:accessorykradial_rn}, which for small $\tilde{r}_{+}$ yields
\begin{equation}\label{eq:lhs_rn}
\begin{split}
\mathrm{lhs} = -\frac{1}{2} &- \frac{1}{4}\biggl[\Delta(\Delta - 4) - \frac{6q^{2}}{\left(1-q^{2}\right)}\\
&- \frac{\tilde{\omega}^{2} - \left(3\tilde{\omega} - 4q\tilde{e}\right)q^{2}\tilde{\omega} - \left(1 + q^{2}\right)q^{2}\tilde{e}^{2} - 2\left(1 - q^{6}\right)}{\left(1-q^{2}\right)^{2}}\biggr]\tilde{r}_{+}^{2} + \mathcal{O}\left(\tilde{r}_{+}^{4}\right)
\end{split}
\end{equation}
while the right hand side organizes as follows
\begin{equation}\label{eq:rhs_rn}
\begin{split}
\mathrm{rhs} = -\frac{1}{2} + \nu_{0}\tilde{r}_{+}^{2} &- \frac{1}{8}\left(1 - q^{2}\right)\left((\Delta - 2)^{2} - \tilde{\omega}^{2}\right)\tilde{r}_{+}^{2} -\frac{\left((1 + q^{6})\tilde{\omega}^{2} - 2(1 + q^{4})q\tilde{e}\tilde{\omega} + (1 + q^{2})q^{2}\tilde{e}^{2}\right)\tilde{r}_{+}^{2}}{4\left(1 - q^{2}\right)^{2}}\\
&- 2\nu_{0}x\left(1 + x + 2x + 5x^{3} + 14x^{4} + \ldots\right)\tilde{r}_{+}^{2} + \mathcal{O}\left(\tilde{r}_{+}^{4}\right),
\end{split}
\end{equation}
where
\begin{equation}\label{eq:sequence_rn}
x = \frac{\left((\Delta - 2)^{2} - \tilde{\omega}^{2}\right)\left(\left(1 + q^{2}\right)\left(\tilde{\omega} - q\tilde{e}\right)^{2} + q^{4}\left(\tilde{\omega}^{2} - \tilde{e}^{2}\right)\right)}{2^{6}\nu_{0}^{2}},
\end{equation}
and the terms inside the parenthesis in equation \eqref{eq:rhs_rn} are associated with the sequence of Catalan numbers \eqref{eq:Catalan}, as presented in the previous section. Equating \eqref{eq:lhs_rn} and \eqref{eq:rhs_rn} leads to an equation for $\nu_{0}$ that can be solved to yield
\begin{equation}\label{eq:nu0_rnads}
\nu_{0} = \pm \frac{1}{8}\sqrt{\begin{aligned}
4\left((\Delta - 2)^{2} - \tilde{\omega}^{2}\right)&\left(\left(1 + q^{2}\right)\left(\tilde{\omega} - q\tilde{e}\right)^{2} + q^{4}\left(\tilde{\omega}^{2} - \tilde{e}^{2}\right)\right)\\
&+ \left(\left(1 + q^{2}\right)\left(3\tilde{\omega}^{2} - \Delta(\Delta - 4)\right) - 4q\tilde{e}\tilde{\omega}\right)^{2}
\end{aligned}},
\end{equation}
which in the limit $q \rightarrow 0$ gives the correction to Schwarzschild-AdS$_{5}$ black-hole solution. Furthermore, $x$ reduces to
\begin{equation}\label{eq:equis_rn}
x = \frac{1}{\left(4  +  \frac{\left(\left(1 + q^{2}\right)\left(3\tilde{\omega}^{2} - \Delta(\Delta - 4)\right) - 4q\tilde{e}\tilde{\omega}\right)^{2}}{\left((\Delta - 2)^{2} - \tilde{\omega}^{2}\right)\left(\left(1 + q^{2}\right)\left(\tilde{\omega} - q\tilde{e}\right)^{2} + q^{4}\left(\tilde{\omega}^{2} - \tilde{e}^{2}\right)\right)}\right)}.
\end{equation}
Analogously, the derivatives of the NS free energy are
\begin{subequations}\label{eq:derivatives_resum_rn}
\begin{equation}\label{eq:dFda_rn}
\partial_{a}F^{\rm inst} = - \left(2x + 3x^{2} + \frac{20}{3}x^{3} + \frac{35}{2}x^{4} + \frac{252}{5}x^{5} + \ldots\right) + \mathcal{O}\left(\tilde{r}_{+}^{2}\right)
\end{equation}
\begin{equation}\label{eq:dFda1_rn}
\begin{split}
\partial_{a_{1}}F^{\rm inst} = \frac{1}{2}(\Delta - 2)\biggl[\left(1 - q^{2}\right) &+ \frac{\left(\left(1 + q^{2}\right)\left(\tilde{\omega} - q\tilde{e}\right)^{2} + q^{4}\left(\tilde{\omega}^{2} - \tilde{e}^{2}\right)\right)}{4\nu_{0}}\\
&\times \left(1 + x + 2x^{2} + 5x^{3} + 14x^{4} + \ldots\right)\biggr]\tilde{r}_{+}^{2} + \mathcal{O}\left(\tilde{r}_{+}^{4}\right)
\end{split}
\end{equation}
\begin{equation}\label{eq:dFdat_rn}
\partial_{a_{t}}F^{\rm inst} = -i\frac{\left(\tilde{\omega} - q\tilde{e}\right)\left((\Delta - 2)^{2} - \tilde{\omega}^{2}\right)}{8\nu_{0}}\left(1 + x + 2x^{2} + 5x^{3} + 14x^{4} + \ldots\right)\tilde{r}_{+} + \mathcal{O}\left(\tilde{r}_{+}^{3}\right)
\end{equation}
\end{subequations}
where $\nu_{0}$ and $x$ are defined in \eqref{eq:nu0_rnads} and \eqref{eq:equis_rn}, respectively. By means of \eqref{eq:sequences}, one can substitute the series expansions in \eqref{eq:derivatives_resum_rn} by 
their generating functions as follows
\begin{subequations}\label{eq:derivatives_of_F_with_x_for_rnads}
\begin{equation}\label{eq:dFda0_rnads}
\partial_{a}F^{\rm inst} \coloneq  \partial_{a}F^{\rm inst}_{(0)} +  \mathcal{O}\left(\tilde{r}_{+}^{2}\right) = \log 4 - 2\log\left(1 + \sqrt{1 - 4x}\right) +  \mathcal{O}\left(\tilde{r}_{+}^{2}\right)
\end{equation}
\begin{equation}\label{eq:dFda12_rnads}
\begin{split}
\partial_{a_{1}}F^{\rm inst} &\coloneq  \partial_{a_{1}}F^{\rm inst}_{(2)}\tilde{r}_{+}^{2} +  \mathcal{O}\left(\tilde{r}_{+}^{4}\right)\\
&= \frac{1}{2}(\Delta - 2)\Biggl[\left(1 - q^{2}\right) + \frac{\left(\left(1 + q^{2}\right)\left(\tilde{\omega} - q\tilde{e}\right)^{2} + q^{4}\left(\tilde{\omega}^{2} - \tilde{e}^{2}\right)\right)}{4\nu_{0}}
\frac{1 - \sqrt{1 - 4x}}{2x}\Biggr]\tilde{r}_{+}^{2} + \mathcal{O}\left(\tilde{r}_{+}^{4}\right)
\end{split}
\end{equation}
\begin{equation}\label{eq:dFdat1_rnads}
\partial_{a_{t}}F^{\rm inst} \coloneq  \partial_{a_{t}}F^{\rm inst}_{(1)}\tilde{r}_{+} +  \mathcal{O}\left(\tilde{r}_{+}^{3}\right) = -i\frac{\left(\tilde{\omega} - q\tilde{e}\right)\left((\Delta - 2)^{2} - \tilde{\omega}^{2}\right)}{8\nu_{0}}\frac{1 - \sqrt{1 - 4x}}{2x}\tilde{r}_{+} + \mathcal{O}\left(\tilde{r}_{+}^{3}\right)
\end{equation}
\end{subequations}
Interestingly, the corrections to the derivatives of $F^{\rm inst}$ in the case of small RN-AdS$_{5}$ BHs match the order of $\tilde{r}_{+}$ in the corrections found in the case of small Kerr-AdS$_{5}$ BHs with equal rotation \eqref{eq:derivatives_of_F_with_x}. Furthermore, the most singular terms at each given order of $\partial_{a}F^{\rm inst}$ are responsible for the appearance of the branch cut after the resummation, as we have seen in the coefficient $\partial_{a}F^{\rm inst}_{(0)}$, which contains a logarithm.

\subsection{Retarded Green's function}
\label{sec:6.1}
The retarded Green's function, defined as the ratio of the response to the source, is given by \eqref{eq:propagator}, where the radial dictionary \eqref{eq:radial_dic} is realized by \eqref{eq:thetas_and_z0_rn}. The new elements in the connection coefficients, such as $a(\ell=0)$ and the derivatives of $F^{\rm inst}$ are obtained from \eqref{eq:a0_rnads} and \eqref{eq:derivatives_of_F_with_x_for_rnads}. Expanding for small $\tilde{r}_{+}$, we derive

\begin{equation}\label{eq:Gret0_rn}
\begin{split}
&G_{\rm ret}\left(\tilde{\omega},\Delta\right) = \frac{e^{-i\pi\Delta}\Gamma\left(2-\Delta\right)\Gamma\left(\half(\Delta-\tilde{\omega}-2)\right)\Gamma\left(\half(\Delta+\tilde{\omega}-2)\right)}{\Gamma\left(\Delta-2\right)\Gamma\left(\half(2-\Delta-\tilde{\omega})\right)\Gamma\left(\half(2-\Delta+\tilde{\omega})\right)}\Biggl\lbrace 1 + \Biggl[\partial_{a_{1}}F^{\rm inst}_{(2)}\\ 
&-\left(\frac{e^{\partial_{a}F^{\rm inst}_{(0)}} - x}{e^{\partial_{a}F^{\rm inst}_{(0)}} + x}\nu_{0} - \frac{1}{4}\left(3(1 + q^{2})\tilde{\omega} - 2q\tilde{e}\right)\right)\left(\psi^{(0)}\left(\thalf(\Delta - \tilde{\omega} - 2)\right) -  \psi^{(0)}\left(\thalf(2 - \Delta - \tilde{\omega})\right)\right)\\
&-\left(\frac{e^{\partial_{a}F^{\rm inst}_{(0)}} - x}{e^{\partial_{a}F^{\rm inst}_{(0)}} + x}\nu_{0} + \frac{1}{4}\left(3(1 + q^{2})\tilde{\omega} - 2q\tilde{e}\right)\right)\left(\psi^{(0)}\left(\thalf(\Delta + \tilde{\omega} - 2)\right) - \psi^{(0)}\left(\thalf(2 - \Delta + \tilde{\omega})\right)\right)\\
&+ \left(\Delta - 2\right)\left(1 + 2q^{2}\right) - \frac{8\nu_{0}(\Delta-2)}{\left((\Delta-2)^{2} - \tilde{\omega}^{2}\right)}\frac{e^{\partial_{a}F^{\rm inst}_{(0)}}}{\left(e^{\partial_{a}F^{\rm inst}_{(0)}} + x\right)}\Biggr]\tilde{r}_{+}^{2}\Biggr\rbrace + \mathcal{O}\left(\tilde{r}_{+}^{3}\right),
\end{split}
\end{equation}
which can be further simplified, yielding
\begin{equation}\label{eq:Gret0_sum_rnads5}
\begin{split}
G_{\rm ret}\left(\tilde{\omega},\Delta\right) = &\frac{e^{-i\pi\Delta}\Gamma\left(2-\Delta\right)\Gamma\left(\half(\Delta-\tilde{\omega}-2)\right)\Gamma\left(\half(\Delta+\tilde{\omega}-2)\right)}{\Gamma\left(\Delta-2\right)\Gamma\left(\half(2-\Delta-\tilde{\omega})\right)\Gamma\left(\half(2-\Delta+\tilde{\omega})\right)}\\
&\Biggl\lbrace 1 + \frac{1}{4}\Biggl[6\left(1 + q^{2}\right)\left(\Delta - 2\right) - \frac{4(\Delta-2)\sqrt{\left((1 + q^{2})(3\tilde{\omega}^{2} - \Delta(\Delta - 4) - 4q\tilde{e}\tilde{\omega}\right)^{2}}}{\left((\Delta-2)^{2} - \tilde{\omega}^{2}\right)}\\
&- \frac{1}{2}\biggl(4q\tilde{e} - 6(1 + q^{2})\tilde{\omega} + \sqrt{\left((1 + q^{2})(3\tilde{\omega}^{2} - \Delta(\Delta - 4) - 4q\tilde{e}\tilde{\omega}\right)^{2}}\biggr)\\
&\times \biggl(\psi^{(0)}\left(\thalf(\Delta - \tilde{\omega} - 2)\right) - \psi^{(0)}\left(\thalf(2 - \Delta - \tilde{\omega})\right)\biggr)\\
&+ \frac{1}{2}\biggl(4q\tilde{e} - 6(1 + q^{2})\tilde{\omega} -  \sqrt{\left((1 + q^{2})(3\tilde{\omega}^{2} - \Delta(\Delta - 4) - 4q\tilde{e}\tilde{\omega}\right)^{2}}\biggr)\\
&\times \biggl(\psi^{(0)}\left(\thalf(\Delta + \tilde{\omega} - 2)\right) - \psi^{(0)}\left(\thalf(2 - \Delta + \tilde{\omega})\right)\biggr)\Biggr]\tilde{r}_{+}^{2}\Biggr\rbrace + \mathcal{O}\left(\tilde{r}_{+}^{3}\right).
\end{split}
\end{equation}

\subsection{Greybody factor}
\label{sec:6.2}
The computation of the greybody factor for the RN-AdS$_{5}$ black hole follows section \ref{sec:5.2}. In the far-region approximation, the radial equation \eqref{eq:radialode_tilde_rn}  simplifies to \eqref{eq:radial_far}, then one can consider that the incoming flux at spatial infinity is 
\begin{equation}\label{eq:flux_in_rn}
\mathcal{F}^{(\infty)}_{\rm in} = - \frac{2 \vert C_{\rm in} \vert^{2} \tilde{\omega}^{4}}{\pi},
\end{equation}
while the flux at the horizon corresponds to the incoming radial solution \eqref{eq:solhor}, expressed in terms of the RN-AdS$_{5}$ black hole parameters:
\begin{equation}\label{eq:flux_hor_rn}
\begin{split}
\mathcal{F}_{\rm hor} = - 2^{\Delta}\vert C_{z_{0}-} \vert^{2} &\left(\tilde{\omega} -q\tilde{e}\sqrt{1 + 2\tilde{r}_{+}^{2}}\right)\tilde{r}_{+}^{3}\\
&\left(\frac{\sqrt{1 + \left(2 + \tilde{r}_{+}^{2} + q^{2}\left(4 + 8\tilde{r}_{+}^{2}\right)\right)\tilde{r}_{+}^{2}}}{1 + 3\tilde{r}_{+}^{2} + \sqrt{1 + \left(2 + \tilde{r}_{+}^{2} + q^{2}\left(4 + 8\tilde{r}_{+}^{2}\right)\right)\tilde{r}_{+}^{2}}}\right)^{\Delta}
\end{split}
\end{equation}
Therefore, the greybody factor for the s-wave reads
\begin{equation}\label{eq:gamma0_rn}
\begin{split}
\gamma^{(\ell = 0)} = \frac{\mathcal{F}_{\rm hor}}{\mathcal{F}^{(\infty)}_{\rm in}} = &\frac{\vert C_{z_{0}-}\vert^{2}}{\vert C_{\rm in} \vert^{2}}\frac{2^{\Delta - 1}\pi \tilde{r}_{+}^{3}}{\tilde{\omega}^{4}}\left(\tilde{\omega} -q\tilde{e}\sqrt{1 + 2\tilde{r}_{+}^{2}}\right)\\
&\qquad \left(\frac{\sqrt{1 + \left(2 + \tilde{r}_{+}^{2} + q^{2}\left(4 + 8\tilde{r}_{+}^{2}\right)\right)\tilde{r}_{+}^{2}}}{1 + 3\tilde{r}_{+}^{2} + \sqrt{1 + \left(2 + \tilde{r}_{+}^{2} + q^{2}\left(4 + 8\tilde{r}_{+}^{2}\right)\right)\tilde{r}_{+}^{2}}}\right)^{\Delta},
\end{split}
\end{equation}
where $C_{\rm in}$ is defined in \eqref{eq:Cin_Cout}. By replacing \eqref{eq:small_a_rn} and \eqref{eq:derivatives_rn} into \eqref{eq:gamma0_rn}, while retaining terms up to second order $\mathcal{O}\left(\tilde{r}_{+}^{2}\right)$ to avoid the pole at $\ell = 0$, the greybody factor in the small $\tilde{r}_{+}$ limit becomes:
\begin{equation}\label{eq:gamma0_nosum_rn}
\begin{split}
\gamma^{(0)} = 2^{2\Delta + 7}\pi^{3}\tilde{r}_{+}^{3}\tilde{\omega}^{2\Delta + 4}\left(\tilde{\omega} - q\tilde{e}\right)&\frac{\sin^{2}\pi\Delta}{\left(\cos\pi\Delta - \cos\pi\tilde{\omega}\right)^{2}}\\
&\textcolor{purple}{\frac{1}{\left(1 + \frac{\left((\Delta - 2)^{2} - \tilde{\omega}^{2}\right)\left(\left(1 + q^{2}\right)\left(\tilde{\omega} - q\tilde{e}\right)^{2} + q^{4}\left(\tilde{\omega}^{2} - \tilde{e}^{2}\right)\right)}{\left(\left(1 + q^{2}\right)\left(3\tilde{\omega}^{2} - \Delta(\Delta - 4)\right) - 4q\tilde{e}\tilde{\omega}\right)^{2}}\right)}}
\frac{1}{\chi} + \mathcal{O}\left(\tilde{r}_{+}^{4}\right)
\end{split}
\end{equation}
where
\begin{equation}
\begin{split}
\chi = 4^{2\Delta}\Gamma\left(\thalf(\Delta - \tilde{\omega} - 2)\right)^{2}&\Gamma\left(\thalf(\Delta + \tilde{\omega} - 2)\right)^{2}\tilde{\omega}^{8}
+ 4^{4}\Gamma\left(\thalf(2 - \Delta - \tilde{\omega})\right)^{2}\Gamma\left(\thalf(2 - \Delta + \tilde{\omega})\right)^{2}\tilde{\omega}^{4\Delta}\\
&\qquad+ \frac{4^{\Delta + 4}\pi^{2}\tilde{\omega}^{2\Delta + 4}\cos 2\pi\Delta}{\left(\cos\pi\Delta - \cos\pi\tilde{\omega}\right)\left((\Delta - 2)^{2} - \tilde{\omega}^{2}\right)},
\end{split}
\end{equation}
Nevertheless, the resummation technique gives a contribution to the asymptotic expansion, determined by \eqref{eq:nu0_rnads} and \eqref{eq:derivatives_of_F_with_x_for_rnads}, 
\begin{equation}\label{eq:gamma0_resum_rnads}
\gamma^{(0)} = 2^{2\Delta + 7}\pi^{3}\tilde{r}_{+}^{3}\tilde{\omega}^{2\Delta + 4}\left(\tilde{\omega} - q\tilde{e}\right)\frac{\sin^{2}\pi\Delta}{\left(\cos\pi\Delta - \cos\pi\tilde{\omega}\right)^{2}}\textcolor{RoyalBlue}{\frac{e^{\partial_{a}F^{\rm inst}_{(0)}}}{\left(e^{\partial_{a}F^{\rm inst}_{(0)}} + x\right)^{2}}}
\frac{1}{\chi} + \mathcal{O}\left(\tilde{r}_{+}^{4}\right)
\end{equation}
\begin{equation}\label{eq:gamma0_resum_rnads5}
\gamma^{(0)} = 2^{2\Delta + 7}\pi^{3}\tilde{r}_{+}^{3}\tilde{\omega}^{2\Delta + 4}\left(\tilde{\omega} - q\tilde{e}\right)\frac{\sin^{2}\pi\Delta}{\left(\cos\pi\Delta - \cos\pi\tilde{\omega}\right)^{2}}\frac{1}{\chi} + \mathcal{O}\left(\tilde{r}_{+}^{4}\right)
\end{equation}

\begin{figure}[h]
\centering
\includegraphics[width=0.6\linewidth]{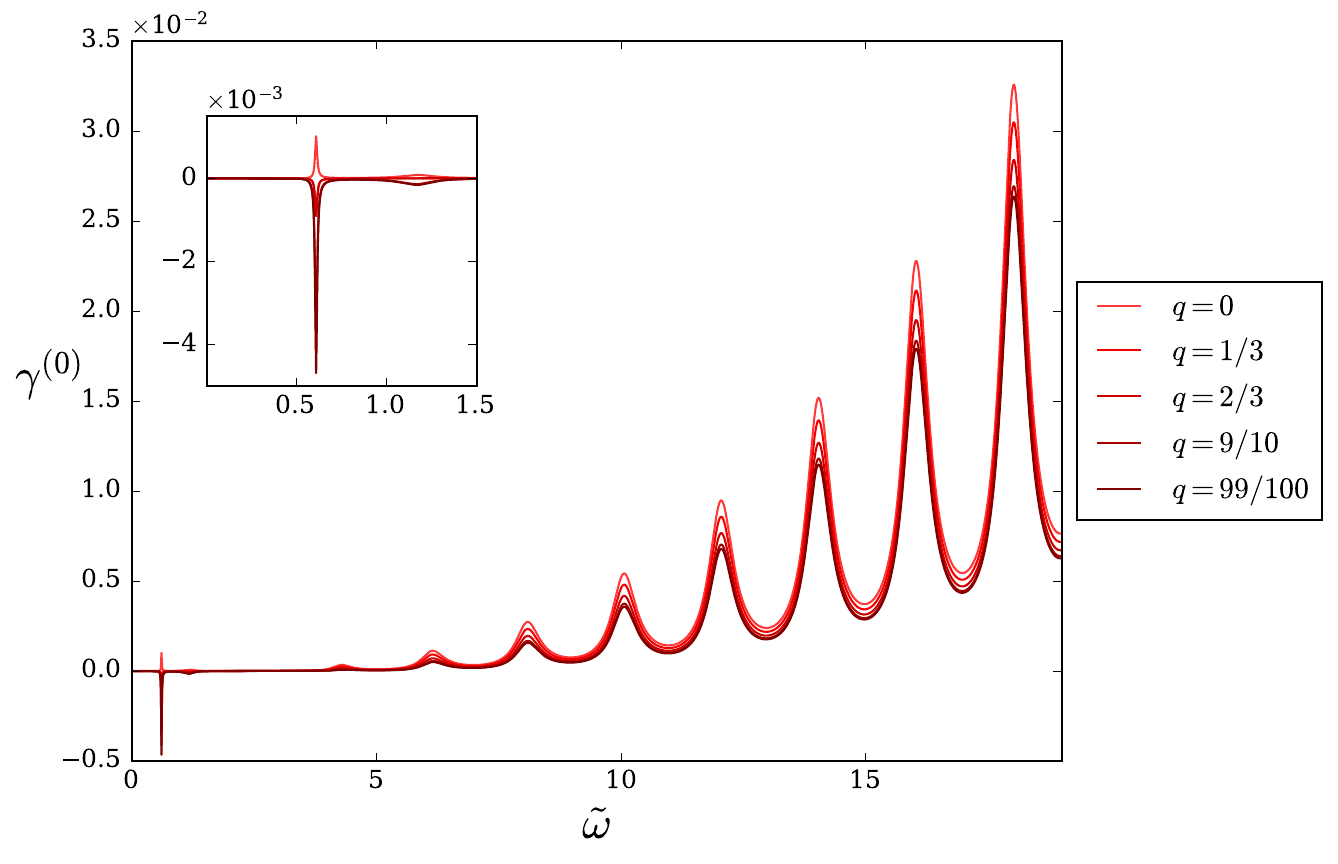}
\caption{Greybody factor \eqref{eq:gamma0_resum_rnads5} as a function of $\tilde{\omega}$ for different values of $q$, and fixed $\tilde{r}_{+}=1/100$, $\tilde{e} = 2\sqrt{3}$, and $\Delta = 46/10$.}
\label{fig:4}
\end{figure}

In Fig. \ref{fig:4}, we present the greybody factor for small RN-AdS$_{5}$ black holes with $\ell =  0$ modes, as calculated using formula \eqref{eq:gamma0_resum_rnads5}. The analysis is performed for different values of the extremality parameter $q = \lbrace 0, 1/3, 2/3, 9/10, 99/100 \rbrace$, while keeping fixed $\tilde{r}_{+} = 1/100$, $\tilde{e} = 2\sqrt{3}$, and $\Delta = 46/10$, and varying the frequency $\tilde{\omega}$. Interestingly, our results show that the greybody spectrum decreases as $q$ increases, which contrasts our previous results in Fig. \ref{fig:1}. Although $\alpha$ and $q$ play the role of extremality parameters, their physical nature is different: the former is associated with the angular momentum of the black hole, while the latter corresponds to the charge of the black hole. It is interesting to note that in the $\gamma^{(0)}$ vs $\tilde{\omega}$ plot for Kerr-AdS$_{5}$, $\gamma^{(0)}$ increases with $\alpha$ for a fixed $\tilde{\omega}$. On the other hand, for RN-AdS$_{5}$ black hole due the presence of a factor $(\tilde{\omega} - q \tilde{e})$, the greybody factor $\gamma^{(0)}$ decreases with $q$ for a fixed value of $\tilde{\omega}$. We see this distinctive behavior by comparing the plot \ref{fig:1} with plot \ref{fig:4}. In addition, for $0 < \tilde{\omega} < q\tilde{e}$, the greybody factor is negative, as shown in the panel on the upper-left corner of figure \ref{fig:4}, pointing out a superradiant frequency window. The first negative peak increases (in absolute value) as we increase $q$, contrary to the behavior seen for $\tilde{\omega} > q\tilde{e}$, where the spectrum shrinks. 

\begin{figure}[h]
\centering
\includegraphics[width=0.6\linewidth]{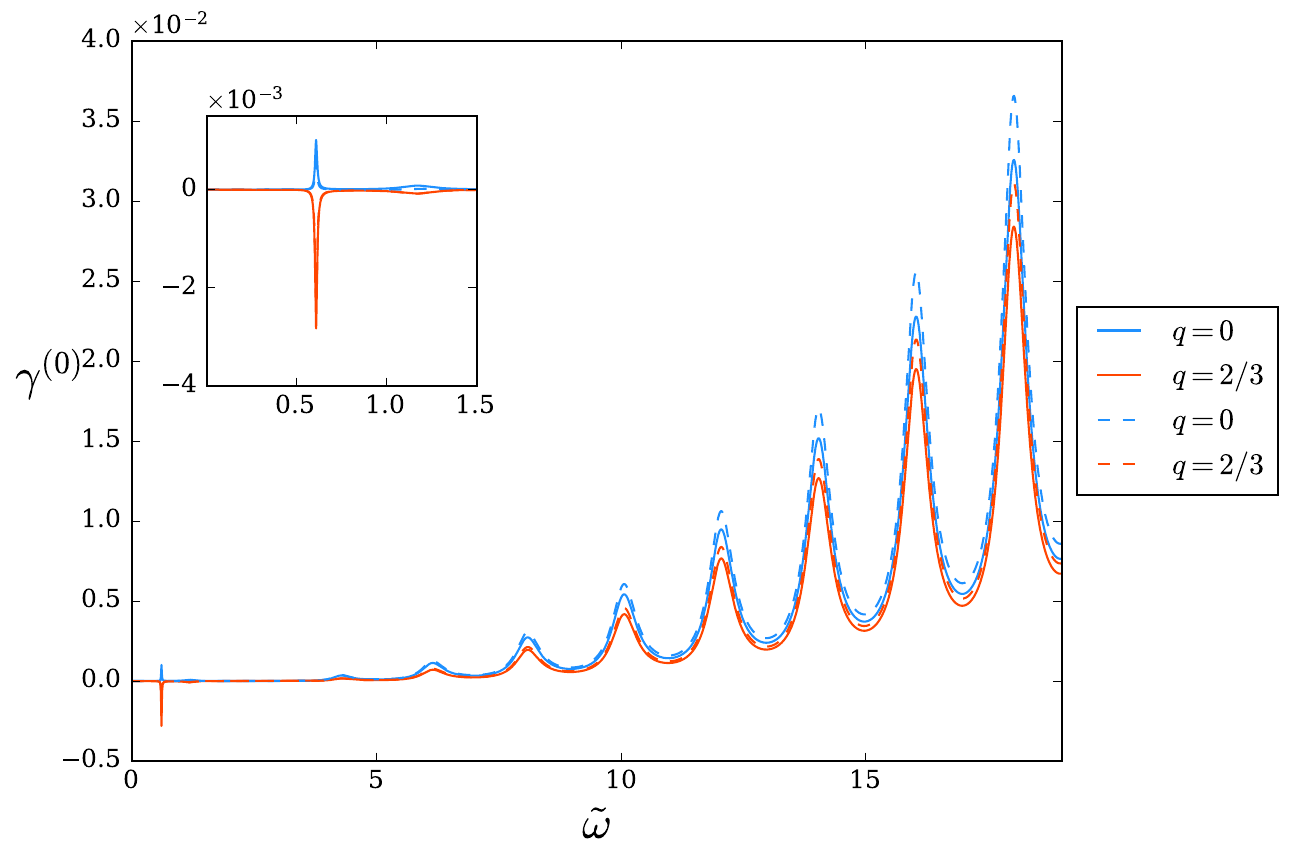}
\caption{Comparison between the greybody factors with resummation (solid lines) and without resummation (dashed lines) for different values of $q$, and fixed $\tilde{r}_{+}=1/100$, $\tilde{e} = 2\sqrt{3}$, and $\Delta = 46/10$.}
\label{fig:5}
\end{figure}

In Fig. \ref{fig:5}, we compare the greybody factor calculated with and without the resummation procedure. The solid curves, representing the asymptotic formula with resummation \eqref{eq:gamma0_resum_rnads5}, show a compressed spectrum compared to the dashed curves, which describe the expression without the resummation \eqref{eq:gamma0_nosum_rn}.

\section{Discussion}
\label{sec:7}
In this paper, we have investigated scalar perturbations in asymptotically AdS sapcetimes, focusing on Kerr-AdS$_{5}$ and Reissner-Nordstr\"{o}m-AdS$_{5}$ black holes. In both cases, the radial ODE can be transformed into a Heun equation --a second-order ODE with four regular singular points. Recently, Heun solutions and their connection coefficients have been computed in terms of Virasoro conformal blocks and, via AGT correspondence, can be expressed in term of the Nekrasov-Shatashvili partition function of an $SU(2)$ supersymmetric gauge theory with four fundamental hypermultiplets \cite{Bonelli:2022ten}. By means of these tools, we have computed the retarded Green's functions and the greybody factor.

Interestingly, these functions introduce new elements, such as the vacuum expectation value of the scalar in the vector hypermultiplet $a$, and the derivatives of the NS function with respect to $a$ and the masses of the gauge theory, $\partial_{a}F^{\rm inst}$ and $\partial_{a_{i}}F^{\rm inst}$, as seen in \eqref{eq:connection_coefs}. We observed that by solving Matone's relation for $a$ at small $\tilde{r}_{+}$, the poles in the analytic expansion of the NS function \eqref{eq:NS_function}, located at $a = \pm n/2$ for $n \in \mathbb{N}$, correspond to poles in the angular momentum quantum number $\ell$ for the series expansion of $a$. Consequently, these poles appear in the expansions of the derivatives of $F^{\rm inst}$. Bearing this in mind, we perform the computation of the retarded Green's function and the greybody factor for the pole at $\ell = 0$ using two strategies: one based on keeping the expansions up to order $\mathcal{O}\left(\tilde{r}_{+}^{2}\right)$ to avoid the pole and one based on curing the pole by considering contributions from all orders in the analytic expansion of the NS function \cite{BarraganAmado:2023apy}. It is worth mentioning that these features are present in both backgrounds, making the computation analogous, and we restrict our analysis for small $\tilde{r}_{+}$ black holes, including an extremality parameter.

At order $\mathcal{O}\left(\tilde{r}_{+}^{2}\right)$, the quantities $a$, $\partial_{a}F^{\rm inst}$, and $\partial_{a_{i}}F^{\rm inst}$ remain finite in the case $\ell = 0$, as shown in \eqref{eq:derivatives_of_F}, such that by substituting these into \eqref{eq:propagator} and expanding for small $\tilde{r}_{+}$, we derive an asymptotic expression for $G_{\rm ret}$ in this limit. For the Kerr-AdS$_{5}$ with equal rotation, the result is given by \eqref{eq:gret0_nosum_knads}, while for RN-AdS$_{5}$, $G_{\rm ret}$ takes the form \eqref{eq:gret0_nosum_rnads}. 

In asymptotically AdS spacetimes, the computation of the greybody factor as the ratio of the flux at the horizon to the incoming flux from infinity is more subtle, due to the asymptotic behavior of the solutions at infinity \eqref{eq:solinf}, which prevents us identifying the ingoing and outgoing wave behavior. However, using a far-region approximation introduced in \cite{Harmark:2007jy}, we can transform the radial equation into a Bessel's diferrential equation \eqref{eq:radial_far}, where the asymptotic form of the solutions reproduces the ingoing and outgoing waves. For the flux at the horizon, we will use the Heun functions and their connections coefficients. Regarding the computation without the resummation procedure, the greybody factors for small Kerr-AdS$_{5}$ and RN-AdS$_{5}$ are given in \eqref{eq:gamma0_nosum_kerrads} and \eqref{eq:gamma0_nosum_rn}, respectively.

On the other hand, the pole structure in the analytic expansion of the instanton part of the NS free energy suggests that at each order, the most singular term provides a contribution that can be resummed. For instance, we have shown that at $\ell = 0$, the first correction $\nu_{0}$ to $a(\ell=0)$ is determined by the generating function of the Catalan numbers \cite{Fitzpatrick:2015foa}. It turns out that the derivatives of $F^{\rm inst}$ also receives corrections from the most singular terms at each order. For instance, the leading singularities from $\partial_{a}F^{\rm inst}$ sum up into a logarithm at order $\mathcal{O}\left(\tilde{r}_{+}^{0}\right)$. Then, we use them into the formulae \eqref{eq:propagator} and \eqref{eq:gamma_ell}. 

The asymptotic expansions for the retarded Green's functions in both spacetimes are corrected in a non-trivial manner, involving terms like $\nu_{0}$, $\partial_{a_{1}}F^{\rm inst}_{(2)}$ and $\tfrac{e^{\partial_{a}F^{\rm inst}_{(0)}} - x}{e^{\partial_{a}F^{\rm inst}_{(0)}} + x}$. Surprisingly, these corrections simplify our results \eqref{eq:Gret0_sum_kerrads5} and \eqref{eq:Gret0_sum_rnads5} in comparison with the asymptotic expansions in Appendix \ref{appendix_C}. 

For the s-wave greybody factor, the effect of resummation simplifies the asymptotic expansions, as well as it leads to a greybody spectrum $\gamma^{(0)}\left(\tilde{\omega}\right)$ that is compressed in comparison with the spectrum without the resummation procedure, as shown in Fig. \ref{fig:3} and \ref{fig:5}. Furthermore, we observe in Fig. \ref{fig:1} that the greybody factor of Kerr-AdS$_{5}$ with equal rotation increases when the rotation parameter $\alpha$ increases, contrary to previous results \cite{Jorge:2014kra}. In Fig. \ref{fig:2} an interesting dynamics show up at small frequencies, as we vary the conformal dimension, two peaks emerge and move closer possibly merge and the separate again. 

For RN-AdS$_{5}$ black holes, the spectrum decreases as we increase the extremality parameter $q$. For $0 < \tilde{\omega} < q\tilde{e}$, the greybody factor is negative, corresponding to a superradiant instability (see Fig. \ref{fig:4}). Finally, it will be interesting to incorporate the higher order corrections in $a$ and see the effect in physical observables. It is also worth to explore the effect of resummation in AdS hyperbolic black hole. 
\section*{Acknowledgments}
We thank Atanu Bhatta for useful discussion and collaboration at the initial stage of the work.
JBA acknowledges financial support from the Funda\c{c}\~{a}o para a Ci\^{e}ncia e a Tecnologia (FCT) - Portugal through the research project 10.54499/2022.03702.PTDC (GENIDE).  SC acknowledges ANRF
grant CRG/2022/006165. AM would like to thank the Council of Scientific and Industrial Research (CSIR), Government of India, for the financial support through a research fellowship (File No.: 09/1005(0034)/2020-EMR-I). 

\appendix
\counterwithin*{equation}{section} 
\renewcommand\theequation{\thesection\arabic{equation}} 

\section{Nekrasov-Shatashvili function with $N_{f}=4$ fundamental hypermultiplets}
\label{appendix_A}
Here, we review the $SU(2)$ Nekrasov-Shatashvili function with $N_f = 4$ fundamental hypermultiplets. Nekrasov considered a deformed lagrangian of $4d$, $\mathcal{N}=4$ gauge theory by introducing two deformation parameters $\epsilon_{1,2}$ \cite{Nekrasov:2002qd,Nekrasov:2003rj}. These parameters parameterize the $SO(4)$ rotation of the spacetime $\mathbb{R}^4$. As a result, the translation symmetry is broken. Nekrasov partition function depends on the coupling parameter $\tau$, VEV $\lq a $' of the adjoint scalars in the vector multiplets, and hypermultiplet
masses $m$. It consists of three parts, namely, the classical, the one-loop and the instanton part:
\begin{equation}
    Z(\tau,a,m;\epsilon_{1,2})= Z_{\text{classical}} Z_{\text{1-loop}} Z_{\text{inst}}
\end{equation}
Its key characteristic is that it provides the prepotential of the theory in the limit $\epsilon_{1,2} \to 0$ and coincides with the prepotential as determined by the Seiberg-Witten curve. Here, we focus on instanton part of Nekrasov partition function 
for $SU(2)$ gauge theory which is obtained from the $U(2)$ partition function by dividing it with $U(1)$-factor \cite{Alday:2009aq,Bonelli:2021uvf}. The $U(2)$ partition function is expressed in terms of the combinatorial formula which we review below.
Let us denote a partition (Young Tableau) by 
\begin{equation}
    Y= (y_1,y_2,...)
\end{equation}
where $y_i$ is the height of the $i$-th column and $y_i=0$, when $i$ is larger than the width of the tableau. Its transpose is given by 
\begin{equation}
    Y^t= (y^t_1,y^t_2,...)
\end{equation}
We write a vector of Young Tableau as
\begin{equation}
    {\bf{Y}}=(Y_1,Y_2)
\end{equation}
For a given Young diagram $Y$, we denote the arm length and the leg length of a box $\lq s$' with respect to the diagram Y as
\begin{equation}
    A_Y(i,j)=y_j-i, \qquad L_Y(i,j)=y^{t}_{i}-j
\end{equation}
where $(i,j)$ denotes the coordinates of the box $\lq s$'. We do not restrict $\lq s$' to be in $Y$. $A_Y$ and  $L_Y$ can be negative if the box $\lq s$' lies outside $Y$ \cite{Aminov:2023jve}.

The instanton part of Nekrasov partition function is given by the summation over the Young tableaux, whose summand is the product of factors corresponding to
the field content of the Lagrangian. 
\begin{equation}
     Z^{U(2)}_{\text{inst}} (t,\vec{a},m;\epsilon_{1,2})=\sum_{\vec{Y}} t^{|\vec{Y}|} z_{\text{vec}}(\vec{a},\vec{Y}) z_{\text{hyp}}(\vec{a},\vec{Y},m),
\end{equation}
where $\vec{a}=(a_1,a_2)$ is the VEV of the scalar in the vector multiplet, $|\vec{Y}|=|Y_1|+|Y_2|$ denotes the total number of boxes (i.e. instanton number) in both $Y_1$ and $Y_2$. The instanton counting parameter $t$ is given by
\begin{equation}
    t=e^{2\pi i\tau }
\end{equation}
where $\tau$ is related with the gauge coupling constant by
\begin{equation}
    \tau=\frac{\theta}{2\pi}+ i \frac{4\pi}{g_{YM}^2}
\end{equation}
The hypermultiplet and vector contributions are given by \cite{Bruzzo:2002xf,Flume:2002az}
\begin{equation}
\begin{split}
    z_{\text{hyp}}(\vec{a},\vec{Y},m) &= \prod_{I=1,2} \prod_{s \in Y_I} \Bigg[a_I+m+\epsilon_1 \left(i-\frac{1}{2}\right)+\epsilon_2 \left(j-\frac{1}{2} \right) \bigg], \\
    z_{\text{vec}}(\vec{a},\vec{Y}) &= \prod_{I,J=1}^2\prod_{s \in Y_I} \frac{1}{a_I-a_J-\epsilon_1 L_{Y_I}(s)+\epsilon_2(A_{Y_I}(s)+1)}\\
    &\qquad\times\prod_{t \in Y_J} \frac{1}{a_I-a_J+\epsilon_1(L_{Y_I}(t)+1)-\epsilon_2 A_{Y_J}(t)}
\end{split}
\end{equation}
In our work, we always consider $a_1=-a_2=a$. Furthermore, let us denote with $m_1,m_2,m_3$ and $m_4$ the masses of the four hypermultiplets. Additionally, we introduce the gauge parameters $a_0,a_t,a_1$ and $a_\infty$ which are related to the masses $m_i$ of
the hypermultiplets via
\begin{equation}\label{eq:masses}
m_{1} = a_{t} + a_{0}, \qquad m_{2} = a_{t} - a_{0}, \qquad m_{3} = a_{1} + a_{\infty}, \qquad m_{4} = a_{1} - a_{\infty}.
\end{equation}
On the other hand, 
the $U(1)$-partition function for $N_f=4$ is given by 
\begin{equation}
    Z_{U(1)}^{N_f=4}=(1-t)^{2(a_1+\frac{\epsilon}{2})(a_t+\frac{\epsilon}{2})/\epsilon_1\epsilon_2}
\end{equation}
where $\epsilon=\epsilon_1+\epsilon_2$. Now, we define  the $SU(2)$ partition function which is given by
\begin{equation}
Z^{SU(2)}_{\text{inst}} (t,\vec{a},m_1,m_2,m_3,m_4;\epsilon_{1,2})= Z^{-1}_{U(1)} (t,m_1,m_2,m_3;\epsilon_{1,2}) Z^{U(2)}_{\text{inst}} (t,\vec{a},m_1,m_2,m_3,m_4;\epsilon_{1,2})
\end{equation}
Moreover, to work in  Nekrasov-Shatashvili (NS) limit, we consider $\epsilon_2 \to 0$ while keeping $\epsilon_1$ fixed (we set $\epsilon_1=1$). Then, the instanton part of NS free energy is defined as 
\begin{equation}
    F_{\rm inst}^{(N_{f}=4)}(a,m_i,t)= \lim_{\epsilon_2 \to 0} \left( \epsilon_2 \log Z^{SU(2)}_{\rm inst} (a,m_i,t;\epsilon_2)\right)
\end{equation}
This can be rewritten as \cite{Bonelli:2021uvf,Dodelson:2022yvn,Aminov:2023jve}
\begin{equation}
    F_{\rm inst}^{(4)}(a,m_i,t)= \lim_{\epsilon_2 \to 0}  \epsilon_2 \log \biggl[ (1-t)^{-2\epsilon_2^{-1}\left(a_1+\frac{1}{2}\right)\left(a_t+\frac{1}{2}\right)} \sum_{\vec{Y}} t^{|\vec{Y}|} z_{\rm vec}(\vec{a},\vec{Y}) z_{\rm hyp}(\vec{a},\vec{Y},m_i) \biggr]
\end{equation}
 An interesting property of NS-free energy is that it is a convergent series in $t$. Hence, it can be written as
 \begin{equation}
      F_{\rm inst}^{(4)}(a,m_i,t)=\sum_{n \ge 1}^\infty c_n(a,m_1,m_2,m_3,m_4) t^n
 \end{equation}
The expansion up to the order $t^2$ is given below;
\begin{equation}\label{eq:NS_function}
\begin{gathered}
F^{(4)}_{\rm inst}\left(a;m_{i};t\right) =  \left[\frac{1}{8}\left(1 - 4a^{2}\right) - \frac{1}{2}\left(m_{1}m_{2} + m_{3}m_{4}\right) - \frac{m_{1}m_{2}m_{3}m_{4}}{2\left(a - \frac{1}{2}\right)} + \frac{m_{1}m_{2}m_{3}m_{4}}{2\left(a + \frac{1}{2}\right)}\right]t \\
+ \biggl[\frac{1}{128}\left(9 - 26a^{2}\right) - \frac{(m_{1} + m_{2})^{2}}{64} - \frac{(m_{3} + m_{4})^{2}}{64} -\frac{7}{32}\left(m_{1}m_{2} + m_{3}m_{4}\right) \\ 
- \frac{\left(1 - 4m_{1}^{2}\right)\left(1 - 4m_{2}^{2}\right)\left(1 - 4m_{3}^{2}\right)\left(1 - 4m_{4}^{2}\right)}{2048\left(a - 1\right)} + \frac{\left(1 - 4m_{1}^{2}\right)\left(1 - 4m_{2}^{2}\right)\left(1 - 4m_{3}^{2}\right)\left(1 - 4m_{4}^{2}\right)}{2048\left(a + 1\right)} \\
- \frac{m_{1}^{2}m_{2}^{2}\left(m_{3}^{2} + m_{4}^{2}\right) + m_{3}^{2}m_{4}^{2}\left(m_{1}^{2} + m_{2}^{2}\right) + 4m_{1}m_{2}m_{3}m_{4}\left(1 - m_{1}m_{2}m_{3}m_{4}\right)}{16\left(a - \frac{1}{2}\right)} \\
+ \frac{m_{1}^{2}m_{2}^{2}\left(m_{3}^{2} + m_{4}^{2}\right) + m_{3}^{2}m_{4}^{2}\left(m_{1}^{2} + m_{2}^{2}\right) + 4m_{1}m_{2}m_{3}m_{4}\left(1 - m_{1}m_{2}m_{3}m_{4}\right)}{16\left(a + \frac{1}{2}\right)} \\
+ \frac{m_{1}^{2}m_{2}^{2}m_{3}^{2}m_{4}^{2}}{16\left(a - \frac{1}{2}\right)^{3}} - \frac{m_{1}^{2}m_{2}^{2}m_{3}^{2}m_{4}^{2}}{16\left(a + \frac{1}{2}\right)^{3}} \biggr]t^{2} + \mathcal{O}\left(t^{3}\right)
\end{gathered}
\end{equation}

\section{Angular Equation}
\label{appendix_B}
By two consecutive transformations $\chi = \sin^{2}\theta$, and $u=\chi/(\chi - \chi_0)$, with $\chi_{0} = (1 - \tilde{a}_{1}^{2})/(\tilde{a}_{2}^{2} - \tilde{a}_{1}^{2})$, we can take the four singular points of \eqref{eq:angularode} to be located at   
\begin{equation}
u=0, \quad u=1, \quad u = u_{0} = \frac{\tilde{a}_{1}^{2} - \tilde{a}_{2}^{2}}{1 - \tilde{a}_2^2}, \quad u = \infty, 
\end{equation}
and the indicial exponents are
\begin{gather}
\alpha^{\pm}_{0} = \pm \frac{m_1}{2},\quad
\alpha^{\pm}_{1} = \frac{1}{2}\left(2 \pm \theta_{\infty}\right),\quad
\alpha^{\pm}_{u_0} = \pm \frac{m_2}{2},\quad
\alpha^{\pm}_{\infty} = \pm\frac{1}{2}\varsigma,
\end{gather}
with
\begin{equation}
\theta_{\infty} = \Delta - 2, \qquad \varsigma = \left(\tilde{\omega} + m_{1}\tilde{a}_{1} + m_{2}\tilde{a}_{2}\right). 
\end{equation}
By introducing the following transformation
\begin{equation}
S(u) = u^{\alpha^{+}_{0}}(u_{0} - u)^{\alpha^{+}_{u_{0}}}(1 - u)^{\alpha^{+}_{1}}y(u),
\label{eq:shomos}
\end{equation}
we bring the angular equation to the canonical Heun form
\begin{equation}
\frac{d^{2}y}{du^2} + \left(\frac{1 + m_{1}}{u} + \frac{1 + m_{2}}{u - u_{0}} + \frac{\Delta - 1}{u - 1}\right)\frac{dy}{du} + \left(\frac{q_{1}q_{2}}{u(u - 1)} - \frac{u_{0}(u_{0} - 1)Q_{0}}{u(u - 1)(u - u_{0})}\right)y(u) = 0,
\label{eq:heunangular}
\end{equation}
where $q_{1},q_{2}$ and the accessory parameter $Q_{0}$ are definend as
\begin{subequations}
\begin{equation}\label{eq:kappasangular}
q_{1} = \frac{1}{2}\left(m_{1} + m_{2} + \Delta - \varsigma\right), \quad\quad q_{2} = \frac{1}{2}\left(m_{1} + m_{2} + \Delta + \varsigma\right),
\end{equation}
\begin{equation}\label{eq:accessoryqangular}
\begin{split}
4u_{0}(u_{0} - 1)Q_{0} &= \frac{\left(\tilde{\omega}^{2} + \tilde{a}_{1}^{2}\Delta(\Delta-4) - \lambda\right)}{1 - \tilde{a}_{2}^{2}} - u_{0}\left[\left(m_{2} + \theta_{\infty} + 1\right)^{2} - m_{2}^{2} - 1\right] \\ 
&- (u_{0} - 1)\left[\left(m_{1} + m_{2} + 1\right)^{2} - \varsigma^{2} - 1\right].
\end{split}
\end{equation}
\end{subequations}

\section{Retarded Green's function}
\label{appendix_C}
For completeness, we provide the asymptotic expansions of the s-wave retarded Green's functions for both the Kerr-AdS$_{5}$ black hole with equal rotation and the RN-AdS$_{5}$ black hole in the small $\tilde{r}_{+}$ limit, without applying the resummation procedure.

\subsubsection*{Kerr-AdS$_{5}$:}
\begin{equation}\label{eq:gret0_nosum_knads}
    \begin{split}
        &G_{\rm ret}\left(\tilde{\omega},\Delta\right) = \frac{e^{-i\pi\Delta}\Gamma\left(2-\Delta\right)\Gamma\left(\half(\Delta-\tilde{\omega}-2)\right)\Gamma\left(\half(\Delta+\tilde{\omega}-2)\right)}{\Gamma\left(\Delta-2\right)\Gamma\left(\half(2-\Delta-\tilde{\omega})\right)\Gamma\left(\half(2-\Delta+\tilde{\omega})\right)}\Biggl\lbrace 1 + \Biggl[\frac{3}{4}\tilde{\omega}\left(1 + \alpha^{2}\right)^{2}\\
        &\times\left(\psi^{(0)}\left(\thalf\left(\Delta - \tilde{\omega} - 2\right)\right) -\psi^{(0)}\left(\thalf\left(2 - \Delta - \tilde{\omega}\right)\right) + \psi^{(0)}\left(\thalf\left(2 - \Delta + \tilde{\omega}\right)\right) - \psi^{(0)}\left(\thalf\left(\Delta +\tilde{\omega} - 2\right)\right)\right)\\
        &- \frac{\left(1 + \alpha^{2}\right)^{2}\left(3\tilde{\omega}^{2} - \Delta(\Delta - 4)\right)(\left(3\tilde{\omega}^{2} - \Delta(\Delta - 4)\right)^{2} - \tilde{\omega}^{2}\left((\Delta - 2)^{2} - \tilde{\omega}^{2}\right))}{8(\left(3\tilde{\omega}^{2} - \Delta(\Delta - 4)\right)^{2} + \tilde{\omega}^{2}\left((\Delta - 2)^{2} - \tilde{\omega}^{2}\right))}\frac{2\pi \sin \pi\Delta}{\left(\cos \pi\Delta - \cos \pi\tilde{\omega}\right)}\\
        &+ \frac{(\Delta - 2)\left(2(1 + \alpha^{4})(\Delta - 3)(\Delta - 1) - \alpha^{2}\left(3\tilde{\omega}^{2} - \Delta(\Delta - 4)\right) - 3(1 + \alpha^{4})\tilde{\omega}^{2}\right)}{\left((\Delta - 2)^{2} - \tilde{\omega}^{2}\right)}\\
        &+ 2\alpha^{2}\left(\Delta - 2\right)\Biggr]\tilde{r}_{+}^{2}\Biggr\rbrace + \mathcal{O}\left(\tilde{r}_{+}^{3}\right)
    \end{split}
\end{equation}

\subsubsection*{RN-AdS$_{5}$:}
\begin{equation}\label{eq:gret0_nosum_rnads}
    \begin{split}
        G_{\rm ret}\left(\tilde{\omega},\Delta\right)&= \frac{e^{-i\pi\Delta}\Gamma\left(2-\Delta\right)\Gamma\left(\half(\Delta-\tilde{\omega}-2)\right)\Gamma\left(\half(\Delta+\tilde{\omega}-2)\right)}{\Gamma\left(\Delta-2\right)\Gamma\left(\half(2-\Delta-\tilde{\omega})\right)\Gamma\left(\half(2-\Delta+\tilde{\omega})\right)}\\
        &\Biggl\lbrace 1 + \Biggl[\frac{\left(3\left(1 + q^{2}\right)\tilde{\omega} - 2q\tilde{e}\right)}{2}\left(\psi^{(0)}\left(\thalf\left(2 - \Delta + \tilde{\omega}\right)\right) - \psi^{(0)}\left(\thalf\left(\Delta + \tilde{\omega} - 2\right)\right)\right)\\
        &+ \frac{\left(\Delta - 2\right)\left((1 + q^{2})(2(\Delta  - 2)^{2} - 3\tilde{\omega}^{2}) - (1 + q^{2})(3\tilde{\omega} + 2) + 2q\tilde{e}(\tilde{\omega} + 1)\right)}{\left((\Delta - 2)^{2} - \tilde{\omega}^{2}\right)}\\
        &+ \Biggl(\frac{\left((1 + q^{2})(3\tilde{\omega}^{2} - \Delta(\Delta - 4)) - 4q\tilde{e}\tilde{\omega}\right)}{8} -\frac{\left(3\left(1 + q^{2}\right)\tilde{\omega} - 2q\tilde{e}\right)}{4}\\
        &+ \frac{1}{4\bigl(\frac{\left((1 + q^{2})(3\tilde{\omega}^{2} - \Delta(\Delta - 4)) - 4q\tilde{e}\tilde{\omega}\right)}{\left(\tilde{\omega}^{2} - (\Delta - 2)^{2}\right)\left((1 +q^{2})(\tilde{\omega} - q\tilde{e})^{2} + q^{4}(\tilde{\omega}^{2} - \tilde{e}^{2})\right)} - \frac{1}{\left((1 + q^{2})(3\tilde{\omega}^{2} - \Delta(\Delta - 4)) - 4q\tilde{e}\tilde{\omega}\right)}\bigr)}\Biggr)\\
        &\biggl(\psi^{(0)}\left(\thalf\left(2 - \Delta - \tilde{\omega}\right)\right) + \psi^{(0)}\left(\thalf\left(2 - \Delta + \tilde{\omega}\right)\right) - \psi^{(0)}\left(\thalf\left(\Delta - \tilde{\omega} - 2\right)\right)\\
        &- \psi^{(0)}\left(\thalf\left(\Delta + \tilde{\omega} - 2\right)\right) - \frac{4(\Delta - 2)}{\left((\Delta - 2)^{2} - \tilde{\omega}^{2}\right)} \biggr)\Biggr]\tilde{r}_{+}^{2}\Biggr\rbrace + \mathcal{O}\left(\tilde{r}_{+}^{3}\right)
    \end{split}
\end{equation}

\bibliography{Resummation}
\bibliographystyle{JHEP}

\end{document}